\documentclass[
 amsmath,amssymb,
 aps,
pra,
twocolumn,
longbibliography,%
]{revtex4-1}
\usepackage[english]{babel}         
\usepackage[latin1]{inputenc}  
\usepackage{amsmath}           
\usepackage{amsfonts}          
\usepackage{amssymb}           
\usepackage{dsfont}
\usepackage{graphicx}

\usepackage{hyperref}
\usepackage{dcolumn}
\usepackage{bm}
\usepackage{color}

\begin{document}
\title{Local Chern marker of smoothly confined Hofstadter fermions}
\author{Urs Gebert}
\author{Bernhard Irsigler}
\author{Walter Hofstetter}
\affiliation{Institut f\"ur Theoretische Physik, Goethe-Universit\"at Frankfurt am Main, Germany}
\begin{abstract} 
The engineering of topological nontrivial states of matter, using cold atoms, has made great progress in the last decade. Driven by experimental successes, it has become of major interest in the cold atom community. In this work we investigate the time-reversal-invariant Hofstadter model with an additional confining potential. By calculating a local spin Chern marker we find that topologically nontrivial phases can be observed in all considered trap geometries. This holds also for spin-orbit coupled fermions, where the model exhibits a quantum spin Hall regime at half filling. Using dynamical mean-field theory, we find that interactions compete against the confining potential and induce a topological phase transition depending on the filling of the system. A further effect of strong interactions yields a magnetic edge, which is localized through the interplay of the density distribution and the underlying topological band structure.

\end{abstract}
\maketitle
\section{Introduction}
Optical lattice experiments offer great possibilities in engineering model Hamiltonians in a clean and well-controlled environment. One focus of current experimental and theoretical interest lies in the investigation of topological states of matter such as the integer or fractional quantum Hall state. Realizations of the paradigmatic Harper-Hofstadter \cite{Aidelsburger2013,Miyake2013} and the Haldane models \cite{Jotzu2014,Flaeschner2016}, which both feature topologically nontrivial states, are showing that these states are now experimentally accessible within cold-atom setups. The topologically nontrivial bulk of a quantum Hall state manifests itself in propagating robust edge states located at the boundary of the system. In cold-atom experiments these boundaries are usually defined by a smooth trapping potential. Recent studies report that this significantly changes the properties of the edge states, as it decreases their group velocity and results in the emerging and splitting of edge states \cite{Tudor2010,Buchhold2012,Goldman2013}. For strong harmonic confinement the trap can even destroy the edge states and therefore the topological phase \cite{Yan2015}. Two-particle interactions, on the other hand, can lead to an enhanced localization of the edge states even in harmonic confinement \cite{Nevado2017,Galilo2017}. The steepness of the trap affects also the bulk of the system and can lead to shrinking \cite{Buchhold2012} and localization \cite{Kolovsky2014} of the bulk. In this work we study the influence of smooth confinement on the topological properties of the bulk and show that these are preserved in different trap geometries. 
Time-reversal (TR) invariant topological insulators can be realized in cold atoms by engineering artificial spin-orbit coupling (SOC) \cite{Galitski2013}. This has been done experimentally in the absence of optical lattices \cite{Lin2011,Wang2012,Cheuk2012,Huang2016} and proposals for the realization of fully tunable TR-invariant SOC in optical lattices exist \cite{Dudarev2004,Grusdt2017}. SOC for bosons on the square lattice leading to a topological nontrivial band structure has also been realized in experiment \cite{Wu2016}.\\
This paper is structured as follows: In Sec. \ref{sec:2} we introduce the underlying Harper-Hofstadter Hamiltonian, including SOC and a staggered potential, and explain the geometry of the additional trap. To analyze the topological properties of our system we use a real-space marker for the Chern number, which we discuss in Sec. \ref{sec:3}. The results are given in section \ref{sec:4}, where we proceed in the following way. In Sec. \ref{sec:4a} we first discuss phase diagrams of a system with noninteracting fermions to pick proper parameters for the calculations including the confining trap potential. We discuss the topological properties for absent SOC (Sec. \ref{sec:4b}) as well as strong SOC (Sec. \ref{sec:4c}) and find topologically nontrivial phases in all trap geometries. Last, we extend our calculations to interacting spin-orbit-coupled fermions in Sec. \ref{sec:4d}. In Sec. \ref{sec:5} we summarize our results.  
\section{Model}\label{sec:2}
The model we consider is the well-known Hofstadter model \cite{Hofstadter1976}, which describes electrons in a two-dimensional square lattice with a strong perpendicular external magnetic field. Here, we use its spinful and TR-invariant
version \cite{Goldman2010}:
\begin{align}
\hat H_0 = -t\sum_{\bm{j}}\left( \hat{\bm{c}}^\dag_{\bm{j}+\hat{x}}e^{i\theta_x}\hat{\bm{c}}_{\bm{j}}  +\hat{\bm{c}}^\dag_{\bm{j}+\hat{y}}e^{i\theta_y}\hat{\bm{c}}_{\bm{j}} + H.c.\right).\label{eq:1}
\end{align}
Here, $\hat{\bm{c}}_{\bm{j}} = (\hat{c}_{\bm{j},\uparrow},\hat{c}_{\bm{j},\downarrow})$ is the annihilation operator on spin-1/2 fermions for lattice site $\bm{j}=(x,y)$, $t$ is the tunneling amplitude, which we set to 1, and $\theta_y = 2\pi \alpha x \sigma_z$  denotes the Peierls phase resulting from  coupling of spins to synthetic gauge fields \cite{Dalibard2011}. It yields an opposite flux per plaquette for different spins. We choose the plaquette flux $\alpha$ to be $1/6$ for our calculations \cite{Orth2013}, which yields a six-band model. $\theta_x=2\pi\gamma\sigma_x$ is a TR-invariant SOC, which mixes different spin states, where we focus on the cases without spin mixing ($\gamma=0$) and with maximal spin mixing ($\gamma=1/4$). Furthermore, we add a staggered potential $\hat{H}_\lambda$, a trap potential $\hat{H}_V$, and in Sec.~\ref{sec:4d} a local Hubbard interaction $\hat{H}_U$ to the Hamiltonian
\begin{align}
\hat{H}= \hat{H}_0+\hat{H}_\lambda+\hat{H}_V+\hat{H}_U,\label{eq:2}
\end{align}
with the staggered potential of amplitude $\lambda$,
\begin{align}
\hat{H}_{\lambda}= \sum_{\bm{j}} (-1)^x \lambda \hat{\bm{c}}_{\bm{j}}^\dag \hat{\bm{c}}_{\bm{j}}.
\end{align}
The staggered potential is used to open a gap at half filling in the case of $\gamma=1/4$ (see Fig. \ref{fig:1})\cite{Orth2013}.
We choose our system to have cylinder geometry, i.e. periodic boundary conditions (PBCs) in the $y$ direction and open boundary conditions (OBCs) with an additional confining trap potential in the $x$ direction,
\begin{align}
\hat{H}_V= \sum_{\bm{j}} V(x) \hat{\bm{c}}_{\bm{j}}^\dag \hat{\bm{c}}_{\bm{j}},\hspace{.55cm} V(x) = V_0\left(\frac{2x}{L-1}-1\right)^\delta,
\end{align}
with parameter $\delta$ to tune the steepness of the trap and $L$ the number of lattice sites in the $x$ direction. $V_0$ is fixed such that the trapping potential has the value $V(0)=V(L-1)=10$ at the boundaries of the system. We investigate the system for a harmonic ($\delta=2$), a quartic ($\delta=4$) and a box-shaped ($\delta \rightarrow \infty$) trap geometry. All calculations are done for a $48\times48$ square lattice. The choice of PBCs in the $y$ direction, instead of OBCs, does, for our system size, not affect the bulk properties we are interested in; however, this cylinder geometry is computationally less demanding. \\
\section{local Chern marker}\label{sec:3}
The topological features of a TR-invariant spin-1/2 system are characterized by a topological $\mathbb{Z}_2$ quantum number \cite{Kane2005,Xu2006,Goldman2010}, which takes the value zero if the system is in a normal insulating (NI) phase and 1 if it is in a quantum spin Hall (QSH) phase. If the spins in the system are not coupled to each other, i.e., $\gamma=0$, the $\mathbb{Z}_2$ index is given by
\begin{align}
\mathbb{Z}_2=\frac{1}{2}(C_{\uparrow} - C_{\downarrow})= C_{\uparrow}\pmod 2,
\end{align} 
where $C_{\sigma}$ is the Chern number \cite{Thouless1982} of the respective spin-$\sigma$ subsystem and the second equality holds only in TR-invariant systems where $C_{\uparrow}=-C_{\downarrow}$. The Chern number is expressed as a Brillouin-zone integral over the Berry curvature and formulated in $\bm{k}$-space. Topological invariants for disordered and interacting systems were defined using the many-body wave function \cite{Niu1985,Sheng2006,Lee2008} or in terms of the single-particle Green's function \cite{Ishikawa1987,Wang2010,Essin2015}. 
However, for strongly inhomogeneous systems a formulation of the Chern number in real space should be more applicable \cite{Prodan2010,Petrescu2012}. One approach, which gives not only a global invariant but yields a spatially resolved quantity to distinguish between topological phases, is the local Chern marker (LCM), developed in Ref.~\cite{Bianco2011} by mapping the $\bm{k}$-space Berry curvature to real space:
\begin{align}
\mathcal{C}(x,y) = -2\pi i \langle x,y|[\hat{P}\hat{x}\hat{P},\hat{P}\hat{y}\hat{P}]|x,y\rangle,
\end{align}
where $\hat{P}$ is the projector onto the occupied states, i.e., onto the states with energies below the Fermi energy $E_F$, and $\hat{x}$ ($\hat{y}$) is the position operator for the $x$ ($y$) direction. This approach was already used to topologically characterize regions in systems with heterojunctions \cite{Bianco2011}, in quasicrystals \cite{Tran2015}, and also in systems with interacting fermions \cite{Amaricci2017,Irsigler2019}. The LCM averaged over the whole lattice, $\langle \mathcal{C} \rangle_{\mathrm{latt}}$, is always zero, since it corresponds to the trace over a commutator. Nevertheless, its average over the bulk area, $\langle \mathcal{C} \rangle_{\mathrm{bulk}}$, gives the expected Chern number with good accuracy, where $\langle \mathcal{C} \rangle_{\mathrm{bulk}}$ is the average over lattice sites with a distance of several sites to the edge of the lattice. The LCM shows nonphysical boundary effects at the edges, to compensate for a finite value in the bulk, which one needs to cut off to get the right bulk average. Referece \cite{Bianco2011} shows that this boundary region broadens if the system is closer to a phase transition. For our calculations a distance of 12 sites turned out to be enough to minimize the error from boundary effects. A scaling analysis of the LCM \cite{Marrazzo2017} has shown that its evaluation over only one unit cell in the center of the system is enough, if the system is sufficiently large, underlining the local character of the LCM. A detailed analysis of the local and nonlocal contributions to the LCM in terms of the single-particle density matrix was done in Ref. \cite{Irsigler2019b}, leading to the conclusion that for systems not to close to the phase transition, the contributions to the local Chern marker are mostly local.  This makes it accessible in experiments by measuring elements of the single-particle density matrix as proposed in \cite{Irsigler2019b,Caio2019} and recently performed in a system of photonic Landau levels \cite{Schine2019}. The non-local contribution to the LCM becomes more important if the system is close to a phase transition, which may be the reason for the broadening of the boundary region mentioned earlier. The size of the boundary region appears to be not dependent on the system size, which we explain by the independence of system size of the contributions to the LCM found in Ref. \cite{Irsigler2019b}.

\section{results}\label{sec:4}
\subsection{Phase diagrams for non interacting fermions}\label{sec:4a}
\begin{figure}
\centering
\includegraphics[width=1\linewidth]{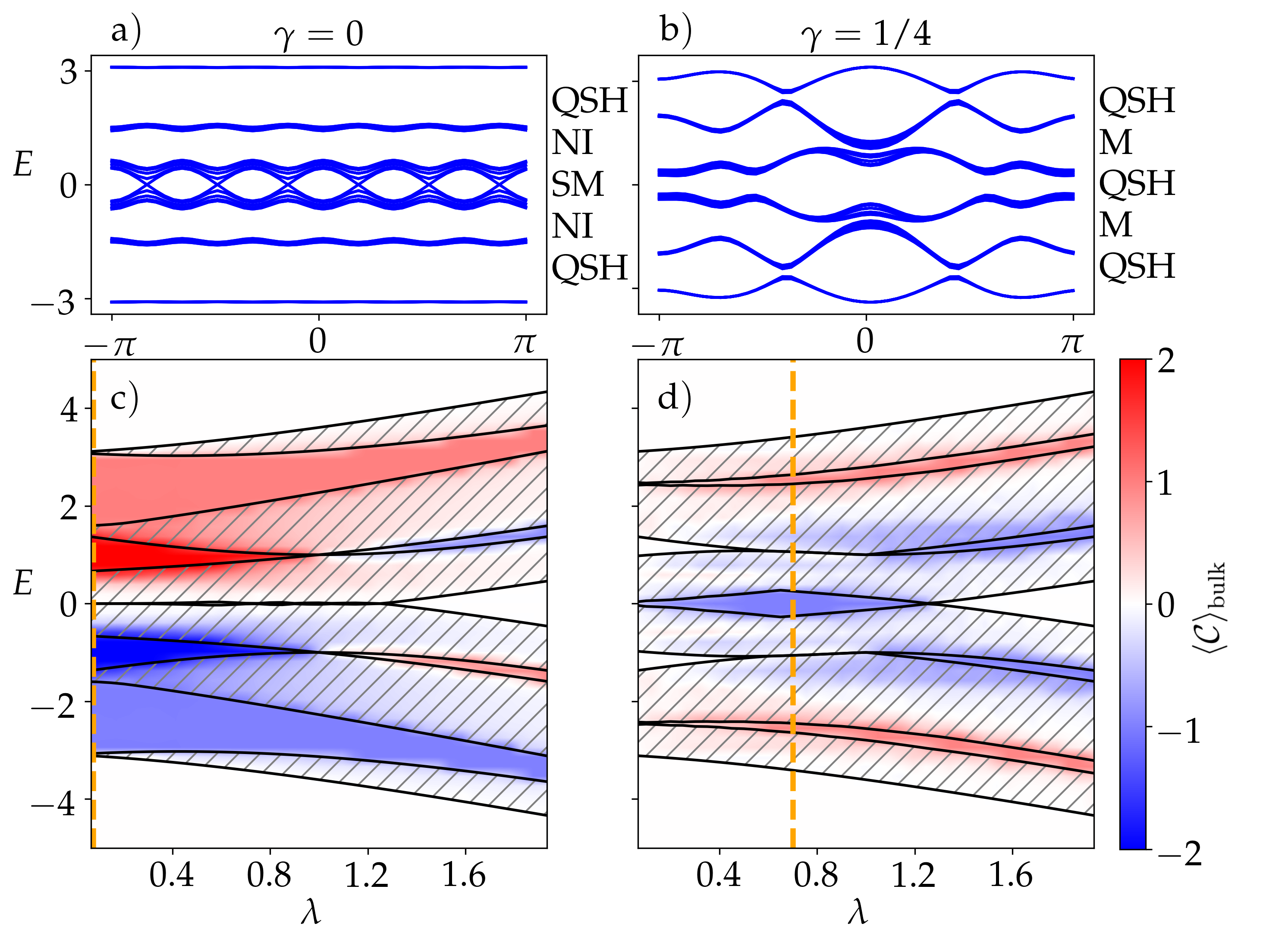}
\caption{Band structure for (a) $\gamma=0$ and $\lambda=0$ and (b) $\gamma=1/4$ and $\lambda= 0.7$, respectively, within PBCs for all directions. Local Chern marker $\mathcal{C}$ calculated for a parameter range of energy $E$ (in units of $t$) and staggered potential $\lambda$, for (c) $\gamma=0$ and (d) $\gamma=1/4$, respectively, corresponding to vanishing and maximal spin-orbit coupling. Hatched areas indicate bulk bands obtained within PBCs. Orange dashed lines correspond to the parameters for the band structures in (a) and (b). Whereas without spin mixing the system is semimetallic (SM) at $E=0$, spin mixing opens a topologically nontrivial gap for values of $\lambda<1.5$ [see (d)].}\label{fig:1}
\end{figure}

We use the LCM to obtain the phase diagrams for noninteracting spin-1/2 fermions as a function of the parameter $\lambda$. The LCM is calculated as described and shown in \hyperref[fig:1]{Figs.~1(c)} and \hyperref[fig:1]{1(d)} for a system with OBCs corresponding to a hard-wall box potential, which gives the same phase diagram as with PBCs \cite{Cocks2012}, since the topological properties of the bulk are not affected by the choice of the boundary conditions. In \hyperref[fig:1]{Figs.~1(a)} and \hyperref[fig:1]{1(b)} we show the corresponding band structure within PBCs for $\lambda=0$ in the case $\gamma=0$ and $\lambda=0.7$ in the case $\gamma=1/4$, which are also the parameters we study in smooth confinement. For the latter the system is gapped at half filling, where between the other bands the gap is rather small or vanishes through indirect band touching leaving the system in a metallic (M) phase.  The LCM shows a finite value also for energies within the bulk bands indicated by hatched areas in \hyperref[fig:1]{Figs.~1(c)} and \hyperref[fig:1]{1(d)}. However, this should not be overinterpreted as the Chern number is only well defined if the Fermi energy lies within a gap. The results of the LCM should therefore also only be valid if the bulk is gapped.
\subsection{Trapped fermions without spin mixing}\label{sec:4b}
\begin{figure}
\centering
\includegraphics[width=1\linewidth]{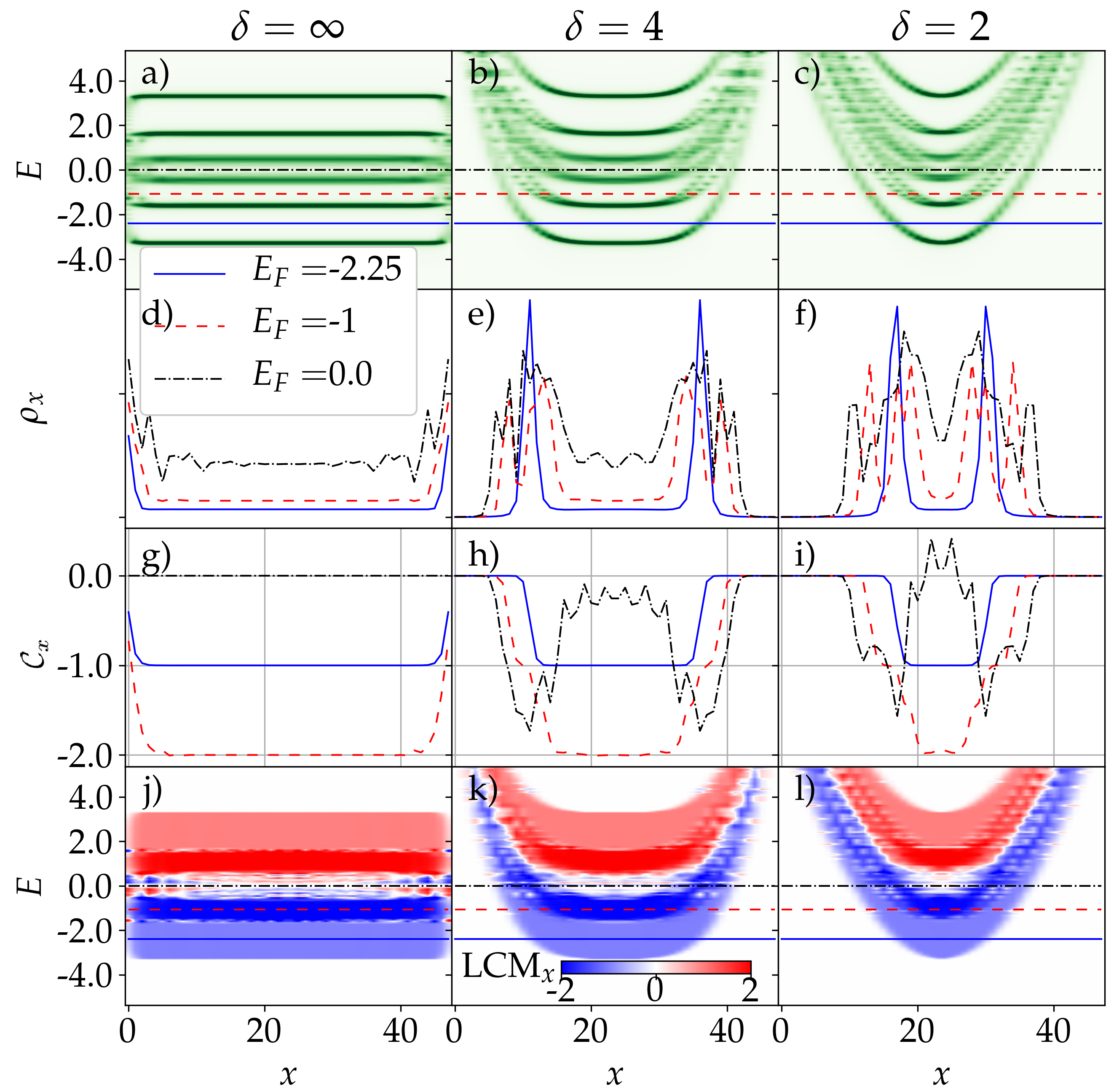}

\caption{(a)-(f) Spectral density and (g)-(l) LCM $\mathcal{C}$ for different traps (hard wall, harmonic, quartic) and vanishing staggered potential and $\gamma=0$. (a)-(c) are Contour plots of the spectral density $\rho_x$ (a.u., $\epsilon=0.05$) evaluated for Fermi energies  $E_F\in [-5,5]$ (in units of $t$). The lines in (d)-(f) show the spectral density for three chosen values of the Fermi energy and correspond to the similar lines in the contour plots (a)-(c). In the same way, (g)-(i) correspond to the lines in the contour plots (j)-(l) of the LCM. A QSH phase, i.e., a LCM of $\pm 1$, is visible within the highest and lowest gap for all three traps. By comparing panels (d)-(f) and (g)-(i), one can see that a change of the LCM is correlated with a peak in the spectral density, i.e., with an edge state.}\label{fig:2}
\end{figure}
We first discuss the case without spin mixing ($\gamma=0$). Here the Hamiltonian $\hat{H}$ is diagonal in spin space and we can directly apply the LCM. The phase diagram shows a QSH phase for a Fermi energy within the lowest and highest gap for all values of $\lambda$. At half filling the gap is closed for $\lambda\lesssim 1.5$ and opens to a NI phase for $\lambda \gtrsim 1.5$. We chose $\lambda=0$ to study the influence of different trap potentials in the system. For this purpose we average the LCM  in the translationally invariant $y$-direction and refer to it as $\mathcal{C}_x$. In Fig.~\ref{fig:2} we compare the results for the  LCM to the momentum integrated spectral density \cite{Buchhold2012}
\begin{align}
\rho^\sigma_x(E,k_y)&= -2\mathrm{Im}\left.\langle k_y|\frac{1}{E+\mu-\hat{H}+i\epsilon - \Sigma(E)}|k_y\rangle\right|^{\sigma\sigma}_{xx}, \\
\rho_x^\sigma(E)&=\int dk_y \rho_x^\sigma(E,k_y),
\end{align}
for a hard wall ($\delta=\infty$), quartic ($\delta=4$) and harmonic confined ($\delta=2$) system. Here $\epsilon\ll 1$ denotes the broadening of the spectral density and $\Sigma(E)$ is the self-energy which vanishes in the noninteracting case. The spectral density shown in  \hyperref[fig:2]{Figs.~2(a)-2(c)} shows six bulk bands, but gapless edge states at the boundary of the system. For a Fermi energy $E_F=0$ the system is semimetallic due to touching bands. We compare the LCM in \hyperref[fig:2]{Figs.~2(g)-2(i)} to the spectral density in \hyperref[fig:2]{Figs.~2(d)-2(f)} for Fermi energies within the two lowest gaps and $E_F=0$. If the Fermi energy lies within a gap, the LCM shows a continuous transition from zero outside the trap to a plateau with constant value in the bulk region. The system is in a QSH phase if only the lowest band is filled, and in a NI phase if two bands are filled. If we compare the behavior of the LCM to the momentum integrated spectral density, we can clearly identify changes in the LCM with peaks in the spectral density. Let us consider for example the blue line in \hyperref[fig:2]{Figs.~2(e)} and \hyperref[fig:2]{2(h)}, which corresponds to a Fermi energy of $E_F=-2.25$. The spectral density shows two large peaks located at the position where the LCM changes from zero to 1 and back. In between these two edge states the system is gapped and in a QSH phase. Since we have spinful fermions, each peak in the spectral density corresponds to a counterpropagating pair of edge states and we see, as stated by the bulk-boundary correspondence, how such a counterpropagating pair of edge states connects topologically distinct regions. The LCM allows us to distinguish these regions in real space even without looking at edge states.  If the Fermi energy lies within the second lowest gap we can see two pairs of edge states in the spectral density and the LCM takes a value of 2 in the middle of the trap. The two edge states can scatter on each other and the system becomes NI. For the gapless regime at half filling the system is also topologically trivial as expected. In \hyperref[fig:2]{Figs.~2(j)-2(l)} we plot the LCM for a large range of Fermi energies. Each horizontal line shows the values of the LCM along the $x$ direction for the corresponding Fermi energy. We can see that the different topological regimes of the system can be found in all trap geometries and that sharp boundaries are not necessarily needed for the realization of a topological insulator.
\subsection{Trapped fermions with spin mixing }\label{sec:4c}
\begin{figure}
\centering
\includegraphics[width=1\linewidth]{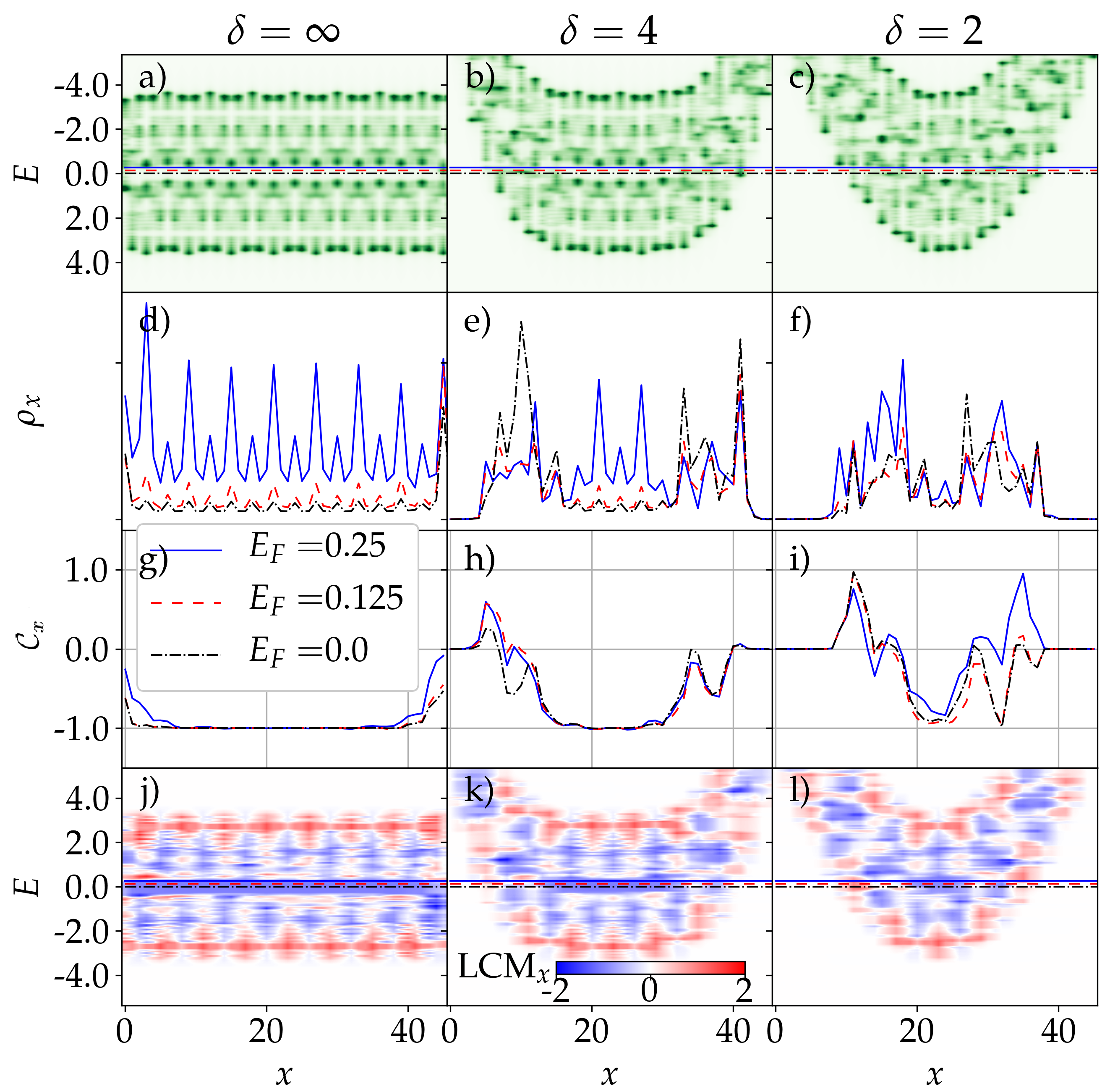}
\caption{(a)-(f) Spectral density $\rho_x$ (a.u., $\epsilon=0.05$) and (g)-(l) LCM  $\mathcal{C}$ for different traps (hard wall, harmonic, quartic) in the case of maximal spin mixing ($\gamma=1/4$, $E$ in units of $t$). (a)-(c) Contour plots of $\rho_x$, where the colored (dashed) lines correspond to (d)-(f). In the same way, the plots in (g)-(i) correspond to lines in the contour plots of the LCM in (j)-(l). The staggered potential is tuned so that the gap at $E_F=0$ has its maximum size. We set $\lambda=0.7$. A quantization of the LCM is visible in all traps. }\label{fig:3}
\end{figure}
Next we consider the case of maximal spin mixing ($\gamma=1/4$). For this case the spin-mixing term in Hamiltonian \eqref{eq:1} simplifies to a hopping in the $x$ direction followed by a spin flip. To apply the LCM the Hamiltonian needs to be decoupled in spin space, which can be achieved by using the pseudospin basis
\begin{align}
\hat{d}_{x,y,\sigma}=\begin{cases} \hat{c}_{x,y,\sigma} & \text{if } x \text{ is even}\\
\hat{c}_{x,y,\bar{\sigma}} & \text{if } x \text{ is odd} 
\end{cases},\label{eq:3}
\end{align} 
where the  Hamiltonian now reads
\begin{align}
\hat{H}_0 = -t\sum_{\bm{j}} \left(\hat{\bm{d}}^\dag_{\bm{j}+\hat{x}} \hat{\bm{d}}_{\bm{j}} +\hat{\bm{d}}^\dag_{\bm{j}+\hat{y}}e^{(-1)^x 2\pi i x \sigma_z \alpha} \hat{\bm{d}}_{\bm{j}} \right)+ \mathrm{H.c.},
\end{align}
with spinors $\hat{\bm{d}}_{\bm{j}}=(\hat{d}_{\bm{j},\uparrow},\hat{d}_{\bm{j},\downarrow})$.
The other terms $\hat{H}_\lambda$ and $\hat{H}_V$ in the full Hamiltonian \eqref{eq:2} are invariant under the basis transformation \eqref{eq:3}. The phase diagram [see \hyperref[fig:1]{Fig.~1(c)}] shows a topological nontrivial gap at half filling, for appropriate values of the staggered potential. In this regime interaction effects are most pronounced, which makes it interesting for further studies on interacting fermions that are discussed in Sec.~\ref{sec:4d}. Therefore, we concentrate on half filling  and set $\lambda=0.7$, where the gap is maximal. Figure \ref{fig:3} shows the results for the spectral density and the LCM. There we average the LCM also over the unit cell in the $x$ direction. The spectral density is very spiky over the whole energy range, but for all trap geometries a gap is visible at $E=0$. Figures \hyperref[fig:3]{3(g)-3(i)} show how the LCM behaves within this gap. For hard-wall and quartic confinement the LCM takes the value of 1 even for Fermi energies close to the bulk band. The LCM shows also a plateau for harmonic confinement, although it is less smooth. Nevertheless, the LCM is approximately 1 in the center of the trap and we could therefore expect the bulk of the system to be in a QSH phase. The relatively small size of the gapped area in harmonic confinement can also lead to finite size effects as the edge states may have a finite overlap. The LCM shows only valid results within a gap, since it is ill defined otherwise.

\subsection{Trapped interacting fermions}\label{sec:4d}

We now study the effect of finite Hubbard interactions; i.e., we add the following local term to the Hamiltonian in Eq.~\eqref{eq:2}:
\begin{equation}
H_U = U\sum_{\bm{j}}\hat{c}_{\bm{j},\uparrow}^\dag\hat{c}_{\bm{j},\uparrow}\hat{c}_{\bm{j},\downarrow}^\dag\hat{c}_{\bm{j},\downarrow},
\end{equation}
where $U$ is the interaction strength. The unconfined as well as the hard-wall confined Hofstadter-Hubbard models have been intensively studied \cite{Cocks2012,Orth2013,Kumar2016,Irsigler2019,Irsigler2019a} in many different aspects; however, the interplay of interactions and smooth confinement is still lacking. We restrict ourselves to the harmonically trapped case, since it is experimentally the most common one and yields a topologically nontrivial state as discussed in the previous sections.

In order to prevent the system from entering a topologically trivial magnetic phase, we adjust the staggered potential as 
\begin{equation}
\lambda=1/2+U/3
\label{staggering}
\end{equation}
according to the phase diagram in Ref.~\cite{Kumar2016}. This ensures a nontrivial bulk topological phase in the center of the trap. For solving the many-body problem, we make use of dynamical mean-field theory (DMFT), using a local self-energy \cite{Georges1996}. Since the systems in our context are highly inhomogeneous, we use the real-space version of DMFT \cite{Okamoto2004,Helmes2008,Snoek2008}. Within real-space DMFT, the full many-body problem on the lattice of $L$ sites is transformed to $L$ single-impurity problems due to the local self-energy $\Sigma^{\sigma\sigma'}_{\bm{i}\bm{j}}(\omega)=\Sigma^{\sigma\sigma'}_{\bm{i}\bm{i}}(\omega)\delta_{\bm{i}\bm{j}}$, where $\omega$ denotes the quasiparticle energy and $\delta_{\bm{i}\bm{j}}$ the Kronecker delta. Since we consider spin-orbit coupled situations, the self-energy can have off-diagonal terms in spin space; i.e., it can be nonzero for $\sigma\neq\sigma'$. For each single-impurity problem, we use a continuous-time quantum Monte Carlo solver in the auxiliary field expansion \cite{Gull2011}. Using spatial symmetries, the number of local self-energies, which have to be calculated, can be reduced. In our case, only the self-energies of one full row in the $x$ direction have to be computed since the system in cylinder geometry is translationally invariant in the $y$ direction. After all single-impurity problems have been solved, a lattice Green's function $G(\omega)$ is constructed from the local self-energies using the Dyson equation

\begin{equation}
\left[G^{-1}(\omega)\right]^{\sigma\sigma'}_{\bm{i}\bm{j}} = \left[G_0^{-1}(\omega)\right]^{\sigma\sigma'}_{\bm{i}\bm{j}} - \Sigma^{\sigma\sigma'}_{\bm{i}\bm{i}}(\omega)\delta_{\bm{i}\bm{j}},
\end{equation}

where $G_0(\omega)$ is the noninteracting lattice Green's function. The new lattice Green's function $G(\omega)$ defines a new lattice problem which is again solved by reducing it to single-impurity problems.  These DMFT iterations are repeated until changes in the self-energy become sufficiently small and self-consistency is reached.

DMFT is formulated in the grand-canonical ensemble, which makes it difficult to solve problems with a fixed number of particles. However, since the latter is experimentally more feasible, we control the mean number of particles per line in the $x$ direction, $N$, by readjusting the chemical potential in each DMFT iteration during the self-consistency procedure. The number of particles cannot be perfectly fixed due to the uncertainty from the quantum Monte Carlo calculations, which increases with increasing interaction strength due to the auxiliary field expansion. Therefore, we refer later to $\bar{N}$ which is defined as the mean $N$ of different $U=1,.\, .\, .\, ,5$.

\begin{figure}
\includegraphics[width=\columnwidth]{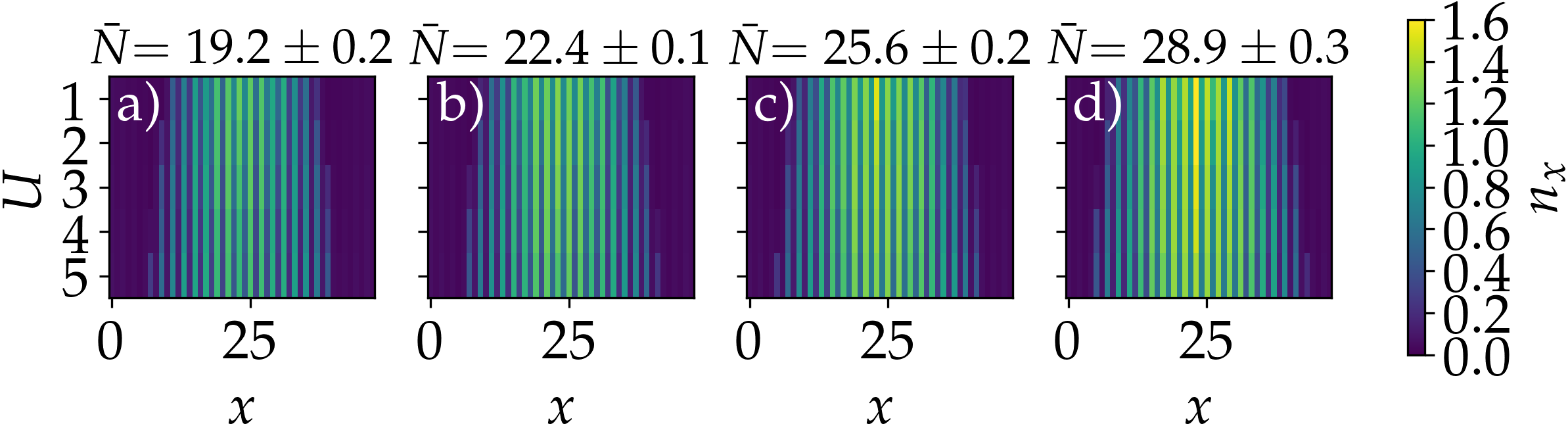}
\caption{Density profiles $n_x$ along a line in the $x$ direction of the harmonic trap as function of interaction strength $U$ for different number of particles, $\bar{N}$. Stronger interactions are broadening the spatial distribution of the particles and decrease the number of doubly occupied sites.}
\label{densities}
\end{figure}

From real-space DMFT we directly get the two-dimensional density profiles $n_{xy}^\sigma$ of spin $\sigma$. Since the system is translationally invariant in the $y$ direction we show the one-dimensional density profiles $n_x = n_x^\uparrow+n_x^\downarrow$ of the harmonically trapped system as a function of the interaction strength for different $\bar{N}$ in Fig.~\ref{densities}. The stripe pattern stems from the staggered potential. We observe that the overall effect of interactions is a spatial broadening of the density profile. Increasing interaction strength also results in a decreasing number of doubly occupied sites. The excess particles are then pushed outwards to occupy sites with higher potential energy of the trap.

\begin{figure}
\includegraphics[width=\columnwidth]{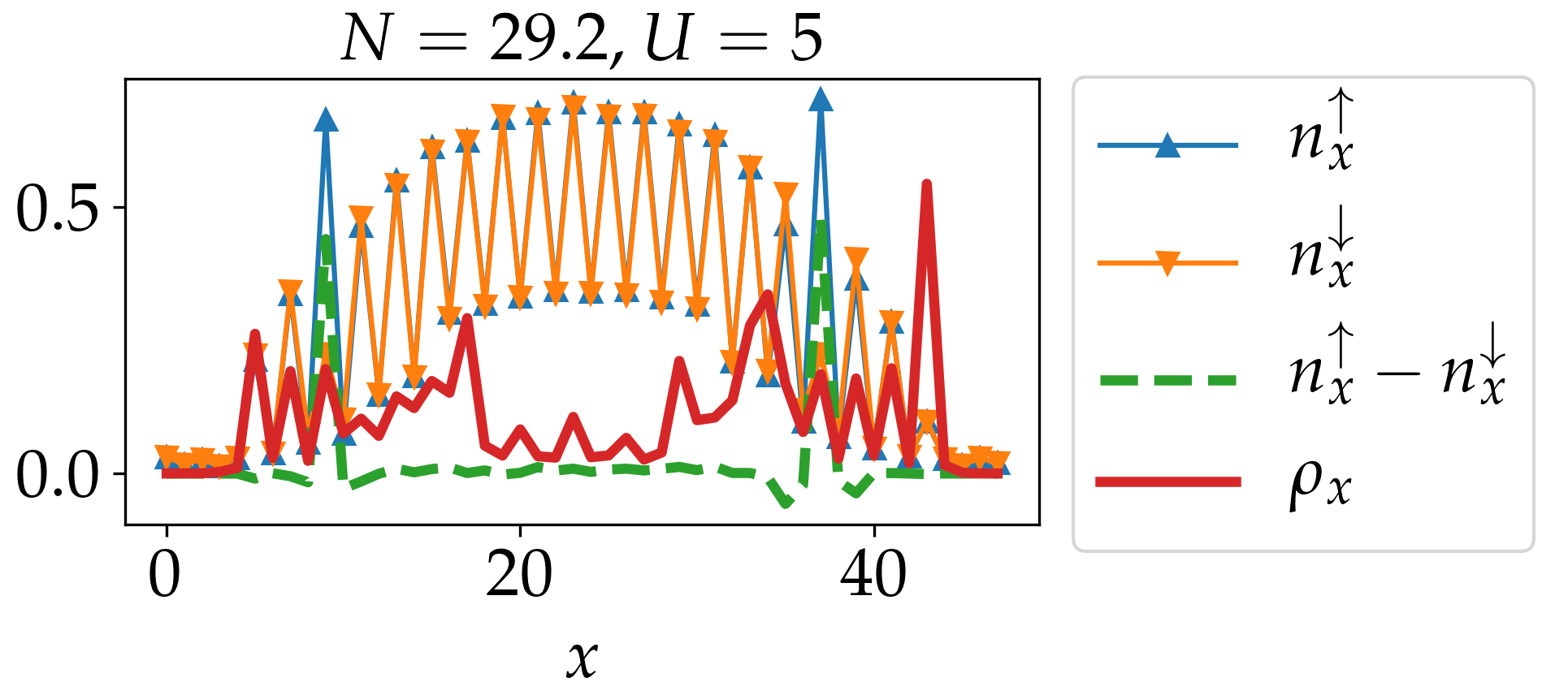}
\caption{Spin-resolved density $n^\sigma_x$ and magnetization profiles $n^\uparrow_x-n^\downarrow_x$ for strong interactions $U$ (in units of $t$). $N$ gives the mean number of particles per line in the $x$ direction. A magnetically strongly localized edge is emerging at the outer region of the trap. The spectral density $\rho_x$ (a.u., $\epsilon=0.01)$ at the Fermi energy shows that the edge state surrounding the bulk QSH phase is further inside and a magnetized phase emerges outside. }
\label{magnetic_edge}
\end{figure}

We observe an interesting effect in the density profiles for strong interactions which is presented in Fig.~\ref{magnetic_edge}. Here, we show exemplarily the spin-resolved density and magnetization profiles for $N=29.2$ and $U=5$. We observe that in the center region, $10<x<35$, as well as in the far edge regions, $x<5$ and $x>40$, the two spin densities are equal and the magnetization vanishes. Locally, however, at $x\approx9$ and $x\approx37$ the occupancy between spin-up and spin-down particles differs and a finite magnetization emerges. Comparing this to the spectral density $\rho_x$ at the Fermi energy shows that the edge states surrounding the bulk QSH phase are further inside. Therefore, we explain this effect in the following way: The bulk QSH phase is protected against magnetization since we control the staggered potential according to Eq.~\eqref{staggering} as explained above. A phase transition to a magnetic phase would only be possible if the gap is closing. Due to the underlying band structure away from the trap center at $x\approx9$ and $x\approx37$, respectively, particles can enter a metallic phase. The gap is thus closed and a magnetic phase can emerge which vanishes again when going even further away from the center where the filling is too small.

\begin{figure}[ht]
\includegraphics[width=\columnwidth]{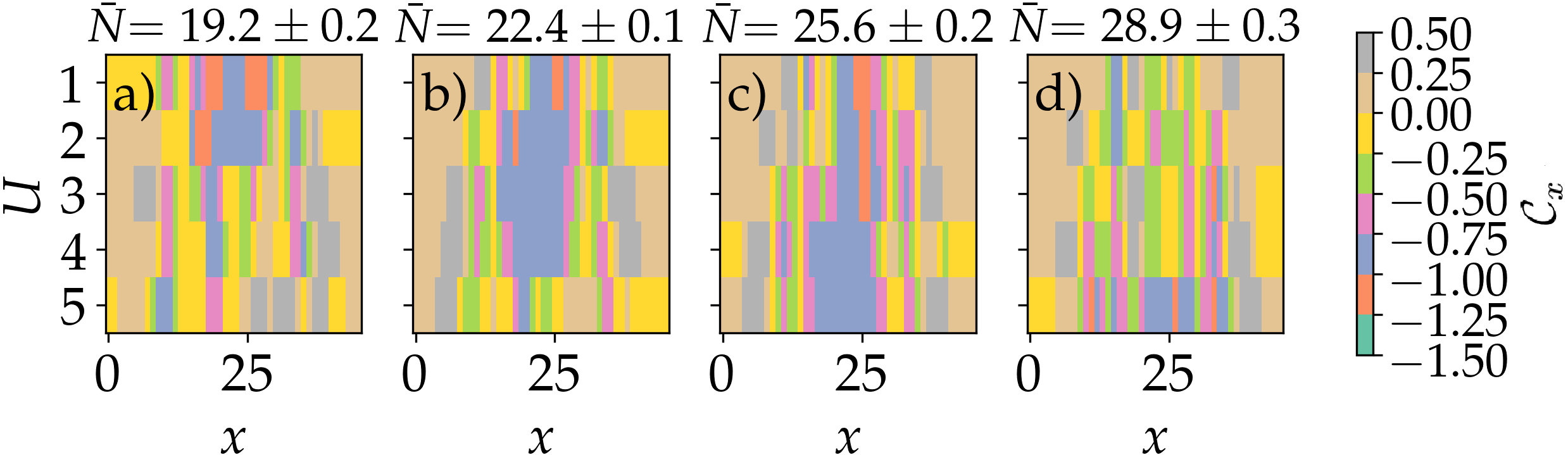}
\caption{Local Chern marker along a line in the $x$ direction, $\mathcal{C}_x$, for the interacting harmonically trapped system as a function of interaction strength $U$ (in units of $t$) for different particle number $\bar{N}$. If there are more particles in the system, i.e., a higher $\bar{N}$, stronger interactions are needed to decrease the filling in the center of the trap and enter the QSH regime.}
\label{interacting_LCM}
\end{figure}

We now turn to the computation of the LCM for the interacting system. To this end, we make use of the topological Hamiltonian approach \cite{Wang2012a}. Here, the interacting Greens function is smoothly transformed to the noninteracting Green's function. If no singularity of this Green's function occurs during this transformation then there is also no gap closing and the topological invariant cannot change. This simplifies computations of topological invariants tremendously. Instead of computing topological invariants of the interacting problem, the topology of the system is covered by an effective, noninteracting Hamiltonian $\left[H_{\text{top}}\right]^{\sigma\sigma'}_{\bm{i}\bm{j}}=\left[H_0\right]^{\sigma\sigma'}_{\bm{i}\bm{j}}+\Sigma^{\sigma\sigma'}_{\bm{i}\bm{i}}(0)\delta_{\bm{i}\bm{j}}$ in matrix representation, where we have used that the self-energy is local. Continuous-time quantum Monte Carlo output is generally expressed in imaginary time which leaves us with the self-energy as function of the fermionic Matsubara frequencies $\omega_n=(2n+1)\pi/\beta$. We determine the zero-frequency self-energy $\Sigma^{\sigma\sigma'}_{\bm{i}\bm{i}}(0)$ by polynomial fitting of $\Sigma^{\sigma\sigma'}_{\bm{i}\bm{i}}(i\omega_n)$ around zero. The combination of the topological Hamiltonian approach and the LCM has been successfully applied in Refs.~\cite{Amaricci2017,Irsigler2019}. We show the interacting LCM in Fig.~\ref{interacting_LCM} as a function of the interaction strength for different $\bar{N}$. Regions with a topologically nontrivial phase, where $\mathcal{C} = -1$, are depicted in blue. Outside these regions the LCM assumes arbitrary values, as we have seen in the noninteracting case in \hyperref[fig:3]{Figs.~3(i)} and \hyperref[fig:3]{3(l)}. We observe that the topologically nontrivial region is shifted to higher interaction strengths as the number of particles in the system is increased. This is due to the fact that interactions push the particles out of the center and then reach half filling in the trap center such that a topologically nontrivial band gap exists. This is a type of interaction-induced topological phase transition \cite{Abanin2012,Vanhala2016,Kumar2016,Zheng2019}; however, here the phase transition is not induced through the competition of interaction strength and staggered potential but rather of interaction strength and trapping potential which completely breaks translational invariance, in contrast to a staggered potential which only increases the size of the unit cell.

\section{conclusion}\label{sec:5}

We apply the local Chern marker to the trapped Hofstadter model and find distinct topologically nontrivial phases even in a smooth confinement. This is complementary to Ref.~\cite{Buchhold2012}. We generalize the treatment to the spin-mixed case, which features a quantum spin Hall gap at half filling. Also here, the local Chern marker indicates topologically nontrivial phases in different trap geometries. In addition, we use dynamical mean-field theory to study the effect of finite on-site interactions. Here, we find an interesting effect of a localized, magnetic edge but nonmagnetic bulk which we explained with topological protection. By using the topological Hamiltonian approach we compute the local Chern marker for the interacting, trapped system and find an interaction-induced topological phase transition depending on the filling.

Based on recent works, we think that our findings can be observed in experiments with tomography methods including a quantum gas microscope. Furthermore, these ideas should be straightforwardly extendable to three dimensions \cite{Scheurer2015} featuring the strong topological insulator phase \cite{Fu2007}.

\begin{acknowledgments}
The authors would like to thank Jun-Hui Zheng for enlightening discussions. This work was supported by the Deutsche Forschungsgemeinschaft (DFG, German Research Foundation) under Project No. 277974659 via Research Unit FOR 2414.  This work was also supported by the Deutsche Forschungsgemeinschaft (DFG) via the high-performance computing center LOEWE-CSC. 
\end{acknowledgments}


\bibliographystyle{apsrev4-1}
\bibliography{library}

\begin{thebibliography}{56}%
\makeatletter
\providecommand \@ifxundefined [1]{%
 \@ifx{#1\undefined}
}%
\providecommand \@ifnum [1]{%
 \ifnum #1\expandafter \@firstoftwo
 \else \expandafter \@secondoftwo
 \fi
}%
\providecommand \@ifx [1]{%
 \ifx #1\expandafter \@firstoftwo
 \else \expandafter \@secondoftwo
 \fi
}%
\providecommand \natexlab [1]{#1}%
\providecommand \enquote  [1]{``#1''}%
\providecommand \bibnamefont  [1]{#1}%
\providecommand \bibfnamefont [1]{#1}%
\providecommand \citenamefont [1]{#1}%
\providecommand \href@noop [0]{\@secondoftwo}%
\providecommand \href [0]{\begingroup \@sanitize@url \@href}%
\providecommand \@href[1]{\@@startlink{#1}\@@href}%
\providecommand \@@href[1]{\endgroup#1\@@endlink}%
\providecommand \@sanitize@url [0]{\catcode `\\12\catcode `\$12\catcode
  `\&12\catcode `\#12\catcode `\^12\catcode `\_12\catcode `\%12\relax}%
\providecommand \@@startlink[1]{}%
\providecommand \@@endlink[0]{}%
\providecommand \url  [0]{\begingroup\@sanitize@url \@url }%
\providecommand \@url [1]{\endgroup\@href {#1}{\urlprefix }}%
\providecommand \urlprefix  [0]{URL }%
\providecommand \Eprint [0]{\href }%
\providecommand \doibase [0]{http://dx.doi.org/}%
\providecommand \selectlanguage [0]{\@gobble}%
\providecommand \bibinfo  [0]{\@secondoftwo}%
\providecommand \bibfield  [0]{\@secondoftwo}%
\providecommand \translation [1]{[#1]}%
\providecommand \BibitemOpen [0]{}%
\providecommand \bibitemStop [0]{}%
\providecommand \bibitemNoStop [0]{.\EOS\space}%
\providecommand \EOS [0]{\spacefactor3000\relax}%
\providecommand \BibitemShut  [1]{\csname bibitem#1\endcsname}%
\let\auto@bib@innerbib\@empty
\bibitem [{\citenamefont {Aidelsburger}\ \emph {et~al.}(2013)\citenamefont
  {Aidelsburger}, \citenamefont {Atala}, \citenamefont {Lohse}, \citenamefont
  {Barreiro}, \citenamefont {Paredes},\ and\ \citenamefont
  {Bloch}}]{Aidelsburger2013}%
  \BibitemOpen
  \bibfield  {author} {\bibinfo {author} {\bibfnamefont {M.}~\bibnamefont
  {Aidelsburger}}, \bibinfo {author} {\bibfnamefont {M.}~\bibnamefont {Atala}},
  \bibinfo {author} {\bibfnamefont {M.}~\bibnamefont {Lohse}}, \bibinfo
  {author} {\bibfnamefont {J.~T.}\ \bibnamefont {Barreiro}}, \bibinfo {author}
  {\bibfnamefont {B.}~\bibnamefont {Paredes}}, \ and\ \bibinfo {author}
  {\bibfnamefont {I.}~\bibnamefont {Bloch}},\ }\href@noop {} {\bibfield
  {journal} {\bibinfo  {journal} {Phys. Rev. Lett.}\ }\textbf {\bibinfo
  {volume} {111}},\ \bibinfo {pages} {185301} (\bibinfo {year}
  {2013})}\BibitemShut {NoStop}%
\bibitem [{\citenamefont {Miyake}\ \emph {et~al.}(2013)\citenamefont {Miyake},
  \citenamefont {Siviloglou}, \citenamefont {Kennedy}, \citenamefont {Burton},\
  and\ \citenamefont {Ketterle}}]{Miyake2013}%
  \BibitemOpen
  \bibfield  {author} {\bibinfo {author} {\bibfnamefont {H.}~\bibnamefont
  {Miyake}}, \bibinfo {author} {\bibfnamefont {G.~A.}\ \bibnamefont
  {Siviloglou}}, \bibinfo {author} {\bibfnamefont {C.~J.}\ \bibnamefont
  {Kennedy}}, \bibinfo {author} {\bibfnamefont {W.~C.}\ \bibnamefont {Burton}},
  \ and\ \bibinfo {author} {\bibfnamefont {W.}~\bibnamefont {Ketterle}},\
  }\href@noop {} {\bibfield  {journal} {\bibinfo  {journal} {Phys. Rev. Lett.}\
  }\textbf {\bibinfo {volume} {111}},\ \bibinfo {pages} {185302} (\bibinfo
  {year} {2013})}\BibitemShut {NoStop}%
\bibitem [{\citenamefont {Jotzu}\ \emph {et~al.}(2014)\citenamefont {Jotzu},
  \citenamefont {Messer}, \citenamefont {Desbuquois}, \citenamefont {Lebrat},
  \citenamefont {Uehlinger}, \citenamefont {Greif},\ and\ \citenamefont
  {Esslinger}}]{Jotzu2014}%
  \BibitemOpen
  \bibfield  {author} {\bibinfo {author} {\bibfnamefont {G.}~\bibnamefont
  {Jotzu}}, \bibinfo {author} {\bibfnamefont {M.}~\bibnamefont {Messer}},
  \bibinfo {author} {\bibfnamefont {R.}~\bibnamefont {Desbuquois}}, \bibinfo
  {author} {\bibfnamefont {M.}~\bibnamefont {Lebrat}}, \bibinfo {author}
  {\bibfnamefont {T.}~\bibnamefont {Uehlinger}}, \bibinfo {author}
  {\bibfnamefont {D.}~\bibnamefont {Greif}}, \ and\ \bibinfo {author}
  {\bibfnamefont {T.}~\bibnamefont {Esslinger}},\ }\href {\doibase
  10.1038/nature13915} {\bibfield  {journal} {\bibinfo  {journal} {Nature}\
  }\textbf {\bibinfo {volume} {515}},\ \bibinfo {pages} {237} (\bibinfo {year}
  {2014})}\BibitemShut {NoStop}%
\bibitem [{\citenamefont {Fl{\"a}schner}\ \emph {et~al.}(2016)\citenamefont
  {Fl{\"a}schner}, \citenamefont {Rem}, \citenamefont {Tarnowski},
  \citenamefont {Vogel}, \citenamefont {L{\"u}hmann}, \citenamefont
  {Sengstock},\ and\ \citenamefont {Weitenberg}}]{Flaeschner2016}%
  \BibitemOpen
  \bibfield  {author} {\bibinfo {author} {\bibfnamefont {N.}~\bibnamefont
  {Fl{\"a}schner}}, \bibinfo {author} {\bibfnamefont {B.~S.}\ \bibnamefont
  {Rem}}, \bibinfo {author} {\bibfnamefont {M.}~\bibnamefont {Tarnowski}},
  \bibinfo {author} {\bibfnamefont {D.}~\bibnamefont {Vogel}}, \bibinfo
  {author} {\bibfnamefont {D.-S.}\ \bibnamefont {L{\"u}hmann}}, \bibinfo
  {author} {\bibfnamefont {K.}~\bibnamefont {Sengstock}}, \ and\ \bibinfo
  {author} {\bibfnamefont {C.}~\bibnamefont {Weitenberg}},\ }\href {\doibase
  10.1126/science.aad4568} {\bibfield  {journal} {\bibinfo  {journal}
  {Science}\ }\textbf {\bibinfo {volume} {352}},\ \bibinfo {pages} {1091}
  (\bibinfo {year} {2016})}\BibitemShut {NoStop}%
\bibitem [{\citenamefont {Stanescu}\ \emph {et~al.}(2010)\citenamefont
  {Stanescu}, \citenamefont {Galitski},\ and\ \citenamefont
  {Das~Sarma}}]{Tudor2010}%
  \BibitemOpen
  \bibfield  {author} {\bibinfo {author} {\bibfnamefont {T.~D.}\ \bibnamefont
  {Stanescu}}, \bibinfo {author} {\bibfnamefont {V.}~\bibnamefont {Galitski}},
  \ and\ \bibinfo {author} {\bibfnamefont {S.}~\bibnamefont {Das~Sarma}},\
  }\href {\doibase 10.1103/PhysRevA.82.013608} {\bibfield  {journal} {\bibinfo
  {journal} {Phys. Rev. A}\ }\textbf {\bibinfo {volume} {82}},\ \bibinfo
  {pages} {013608} (\bibinfo {year} {2010})}\BibitemShut {NoStop}%
\bibitem [{\citenamefont {Buchhold}\ \emph {et~al.}(2012)\citenamefont
  {Buchhold}, \citenamefont {Cocks},\ and\ \citenamefont
  {Hofstetter}}]{Buchhold2012}%
  \BibitemOpen
  \bibfield  {author} {\bibinfo {author} {\bibfnamefont {M.}~\bibnamefont
  {Buchhold}}, \bibinfo {author} {\bibfnamefont {D.}~\bibnamefont {Cocks}}, \
  and\ \bibinfo {author} {\bibfnamefont {W.}~\bibnamefont {Hofstetter}},\
  }\href@noop {} {\bibfield  {journal} {\bibinfo  {journal} {Phys. Rev. A}\
  }\textbf {\bibinfo {volume} {\textbf{85}}},\ \bibinfo {pages} {063614}
  (\bibinfo {year} {2012})}\BibitemShut {NoStop}%
\bibitem [{\citenamefont {Goldman}\ \emph {et~al.}(2013)\citenamefont
  {Goldman}, \citenamefont {Dalibard}, \citenamefont {Dauphin}, \citenamefont
  {Gerbier}, \citenamefont {Lewenstein}, \citenamefont {Zoller},\ and\
  \citenamefont {Spielman}}]{Goldman2013}%
  \BibitemOpen
  \bibfield  {author} {\bibinfo {author} {\bibfnamefont {N.}~\bibnamefont
  {Goldman}}, \bibinfo {author} {\bibfnamefont {J.}~\bibnamefont {Dalibard}},
  \bibinfo {author} {\bibfnamefont {A.}~\bibnamefont {Dauphin}}, \bibinfo
  {author} {\bibfnamefont {F.}~\bibnamefont {Gerbier}}, \bibinfo {author}
  {\bibfnamefont {M.}~\bibnamefont {Lewenstein}}, \bibinfo {author}
  {\bibfnamefont {P.}~\bibnamefont {Zoller}}, \ and\ \bibinfo {author}
  {\bibfnamefont {I.~B.}\ \bibnamefont {Spielman}},\ }\href {\doibase
  10.1073/pnas.1300170110} {\bibfield  {journal} {\bibinfo  {journal} {P. Natl.
  Acad. Sci. USA}\ }\textbf {\bibinfo {volume} {110}},\ \bibinfo {pages} {6736}
  (\bibinfo {year} {2013})}\BibitemShut {NoStop}%
\bibitem [{\citenamefont {Yan}\ \emph {et~al.}(2015)\citenamefont {Yan},
  \citenamefont {Li}, \citenamefont {Yang},\ and\ \citenamefont
  {Wan}}]{Yan2015}%
  \BibitemOpen
  \bibfield  {author} {\bibinfo {author} {\bibfnamefont {Z.}~\bibnamefont
  {Yan}}, \bibinfo {author} {\bibfnamefont {B.}~\bibnamefont {Li}}, \bibinfo
  {author} {\bibfnamefont {X.}~\bibnamefont {Yang}}, \ and\ \bibinfo {author}
  {\bibfnamefont {S.}~\bibnamefont {Wan}},\ }\href
  {https://doi.org/10.1038/srep16197} {\bibfield  {journal} {\bibinfo
  {journal} {Sci. Rep.}\ }\textbf {\bibinfo {volume} {5}},\ \bibinfo {pages}
  {16197} (\bibinfo {year} {2015})}\BibitemShut {NoStop}%
\bibitem [{\citenamefont {Nevado}\ \emph {et~al.}(2017)\citenamefont {Nevado},
  \citenamefont {Fern\'andez-Lorenzo},\ and\ \citenamefont
  {Porras}}]{Nevado2017}%
  \BibitemOpen
  \bibfield  {author} {\bibinfo {author} {\bibfnamefont {P.}~\bibnamefont
  {Nevado}}, \bibinfo {author} {\bibfnamefont {S.}~\bibnamefont
  {Fern\'andez-Lorenzo}}, \ and\ \bibinfo {author} {\bibfnamefont
  {D.}~\bibnamefont {Porras}},\ }\href {\doibase
  10.1103/PhysRevLett.119.210401} {\bibfield  {journal} {\bibinfo  {journal}
  {Phys. Rev. Lett.}\ }\textbf {\bibinfo {volume} {119}},\ \bibinfo {pages}
  {210401} (\bibinfo {year} {2017})}\BibitemShut {NoStop}%
\bibitem [{\citenamefont {Galilo}\ \emph {et~al.}(2017)\citenamefont {Galilo},
  \citenamefont {Lee},\ and\ \citenamefont {Barnett}}]{Galilo2017}%
  \BibitemOpen
  \bibfield  {author} {\bibinfo {author} {\bibfnamefont {B.}~\bibnamefont
  {Galilo}}, \bibinfo {author} {\bibfnamefont {D.~K.~K.}\ \bibnamefont {Lee}},
  \ and\ \bibinfo {author} {\bibfnamefont {R.}~\bibnamefont {Barnett}},\ }\href
  {\doibase 10.1103/PhysRevLett.119.203204} {\bibfield  {journal} {\bibinfo
  {journal} {Phys. Rev. Lett.}\ }\textbf {\bibinfo {volume} {119}},\ \bibinfo
  {pages} {203204} (\bibinfo {year} {2017})}\BibitemShut {NoStop}%
\bibitem [{\citenamefont {Kolovsky}\ \emph {et~al.}(2014)\citenamefont
  {Kolovsky}, \citenamefont {Grusdt},\ and\ \citenamefont
  {Fleischhauer}}]{Kolovsky2014}%
  \BibitemOpen
  \bibfield  {author} {\bibinfo {author} {\bibfnamefont {A.~R.}\ \bibnamefont
  {Kolovsky}}, \bibinfo {author} {\bibfnamefont {F.}~\bibnamefont {Grusdt}}, \
  and\ \bibinfo {author} {\bibfnamefont {M.}~\bibnamefont {Fleischhauer}},\
  }\href {\doibase 10.1103/PhysRevA.89.033607} {\bibfield  {journal} {\bibinfo
  {journal} {Phys. Rev. A}\ }\textbf {\bibinfo {volume} {89}},\ \bibinfo
  {pages} {033607} (\bibinfo {year} {2014})}\BibitemShut {NoStop}%
\bibitem [{\citenamefont {Galitski}\ and\ \citenamefont
  {Spielman}(2013)}]{Galitski2013}%
  \BibitemOpen
  \bibfield  {author} {\bibinfo {author} {\bibfnamefont {V.}~\bibnamefont
  {Galitski}}\ and\ \bibinfo {author} {\bibfnamefont {I.~B.}\ \bibnamefont
  {Spielman}},\ }\href {\doibase 10.1038/nature11841} {\bibfield  {journal}
  {\bibinfo  {journal} {Nature}\ }\textbf {\bibinfo {volume} {494}},\ \bibinfo
  {pages} {49} (\bibinfo {year} {2013})}\BibitemShut {NoStop}%
\bibitem [{\citenamefont {Lin}\ \emph {et~al.}(2011)\citenamefont {Lin},
  \citenamefont {Jim\'{e}nez-Garc\'{i}a},\ and\ \citenamefont
  {Spielman}}]{Lin2011}%
  \BibitemOpen
  \bibfield  {author} {\bibinfo {author} {\bibfnamefont {Y.-J.}\ \bibnamefont
  {Lin}}, \bibinfo {author} {\bibfnamefont {K.}~\bibnamefont
  {Jim\'{e}nez-Garc\'{i}a}}, \ and\ \bibinfo {author} {\bibfnamefont {I.~B.}\
  \bibnamefont {Spielman}},\ }\href {\doibase 10.1038/nature09887} {\bibfield
  {journal} {\bibinfo  {journal} {Nature}\ }\textbf {\bibinfo {volume} {471}},\
  \bibinfo {pages} {83} (\bibinfo {year} {2011})}\BibitemShut {NoStop}%
\bibitem [{\citenamefont {Wang}\ \emph {et~al.}(2012)\citenamefont {Wang},
  \citenamefont {Yu}, \citenamefont {Fu}, \citenamefont {Miao}, \citenamefont
  {Huang}, \citenamefont {Chai}, \citenamefont {Zhai},\ and\ \citenamefont
  {Zhang}}]{Wang2012}%
  \BibitemOpen
  \bibfield  {author} {\bibinfo {author} {\bibfnamefont {P.}~\bibnamefont
  {Wang}}, \bibinfo {author} {\bibfnamefont {Z.-Q.}\ \bibnamefont {Yu}},
  \bibinfo {author} {\bibfnamefont {Z.}~\bibnamefont {Fu}}, \bibinfo {author}
  {\bibfnamefont {J.}~\bibnamefont {Miao}}, \bibinfo {author} {\bibfnamefont
  {L.}~\bibnamefont {Huang}}, \bibinfo {author} {\bibfnamefont
  {S.}~\bibnamefont {Chai}}, \bibinfo {author} {\bibfnamefont {H.}~\bibnamefont
  {Zhai}}, \ and\ \bibinfo {author} {\bibfnamefont {J.}~\bibnamefont {Zhang}},\
  }\href {\doibase 10.1103/PhysRevLett.109.095301} {\bibfield  {journal}
  {\bibinfo  {journal} {Phys. Rev. Lett.}\ }\textbf {\bibinfo {volume} {109}},\
  \bibinfo {pages} {095301} (\bibinfo {year} {2012})}\BibitemShut {NoStop}%
\bibitem [{\citenamefont {Cheuk}\ \emph {et~al.}(2012)\citenamefont {Cheuk},
  \citenamefont {Sommer}, \citenamefont {Hadzibabic}, \citenamefont {Yefsah},
  \citenamefont {Bakr},\ and\ \citenamefont {Zwierlein}}]{Cheuk2012}%
  \BibitemOpen
  \bibfield  {author} {\bibinfo {author} {\bibfnamefont {L.~W.}\ \bibnamefont
  {Cheuk}}, \bibinfo {author} {\bibfnamefont {A.~T.}\ \bibnamefont {Sommer}},
  \bibinfo {author} {\bibfnamefont {Z.}~\bibnamefont {Hadzibabic}}, \bibinfo
  {author} {\bibfnamefont {T.}~\bibnamefont {Yefsah}}, \bibinfo {author}
  {\bibfnamefont {W.~S.}\ \bibnamefont {Bakr}}, \ and\ \bibinfo {author}
  {\bibfnamefont {M.~W.}\ \bibnamefont {Zwierlein}},\ }\href {\doibase
  10.1103/PhysRevLett.109.095302} {\bibfield  {journal} {\bibinfo  {journal}
  {Phys. Rev. Lett.}\ }\textbf {\bibinfo {volume} {109}},\ \bibinfo {pages}
  {095302} (\bibinfo {year} {2012})}\BibitemShut {NoStop}%
\bibitem [{\citenamefont {Huang}\ \emph {et~al.}(2016)\citenamefont {Huang},
  \citenamefont {Meng}, \citenamefont {Wang}, \citenamefont {Peng},
  \citenamefont {Zhang}, \citenamefont {Chen}, \citenamefont {Li},
  \citenamefont {Zhou},\ and\ \citenamefont {Zhang}}]{Huang2016}%
  \BibitemOpen
  \bibfield  {author} {\bibinfo {author} {\bibfnamefont {L.}~\bibnamefont
  {Huang}}, \bibinfo {author} {\bibfnamefont {Z.}~\bibnamefont {Meng}},
  \bibinfo {author} {\bibfnamefont {P.}~\bibnamefont {Wang}}, \bibinfo {author}
  {\bibfnamefont {P.}~\bibnamefont {Peng}}, \bibinfo {author} {\bibfnamefont
  {S.-L.}\ \bibnamefont {Zhang}}, \bibinfo {author} {\bibfnamefont
  {L.}~\bibnamefont {Chen}}, \bibinfo {author} {\bibfnamefont {D.}~\bibnamefont
  {Li}}, \bibinfo {author} {\bibfnamefont {Q.}~\bibnamefont {Zhou}}, \ and\
  \bibinfo {author} {\bibfnamefont {J.}~\bibnamefont {Zhang}},\ }\href
  {\doibase 10.1038/nphys3672} {\bibfield  {journal} {\bibinfo  {journal}
  {Nature Physics}\ }\textbf {\bibinfo {volume} {12}},\ \bibinfo {pages} {540}
  (\bibinfo {year} {2016})}\BibitemShut {NoStop}%
\bibitem [{\citenamefont {Dudarev}\ \emph {et~al.}(2004)\citenamefont
  {Dudarev}, \citenamefont {Diener}, \citenamefont {Carusotto},\ and\
  \citenamefont {Niu}}]{Dudarev2004}%
  \BibitemOpen
  \bibfield  {author} {\bibinfo {author} {\bibfnamefont {A.~M.}\ \bibnamefont
  {Dudarev}}, \bibinfo {author} {\bibfnamefont {R.~B.}\ \bibnamefont {Diener}},
  \bibinfo {author} {\bibfnamefont {I.}~\bibnamefont {Carusotto}}, \ and\
  \bibinfo {author} {\bibfnamefont {Q.}~\bibnamefont {Niu}},\ }\href {\doibase
  10.1103/PhysRevLett.92.153005} {\bibfield  {journal} {\bibinfo  {journal}
  {Phys. Rev. Lett.}\ }\textbf {\bibinfo {volume} {92}},\ \bibinfo {pages}
  {153005} (\bibinfo {year} {2004})}\BibitemShut {NoStop}%
\bibitem [{\citenamefont {Grusdt}\ \emph {et~al.}(2017)\citenamefont {Grusdt},
  \citenamefont {Li}, \citenamefont {Bloch},\ and\ \citenamefont
  {Demler}}]{Grusdt2017}%
  \BibitemOpen
  \bibfield  {author} {\bibinfo {author} {\bibfnamefont {F.}~\bibnamefont
  {Grusdt}}, \bibinfo {author} {\bibfnamefont {T.}~\bibnamefont {Li}}, \bibinfo
  {author} {\bibfnamefont {I.}~\bibnamefont {Bloch}}, \ and\ \bibinfo {author}
  {\bibfnamefont {E.}~\bibnamefont {Demler}},\ }\href {\doibase
  10.1103/PhysRevA.95.063617} {\bibfield  {journal} {\bibinfo  {journal} {Phys.
  Rev. A}\ }\textbf {\bibinfo {volume} {95}},\ \bibinfo {pages} {063617}
  (\bibinfo {year} {2017})}\BibitemShut {NoStop}%
\bibitem [{\citenamefont {Wu}\ \emph {et~al.}(2016)\citenamefont {Wu},
  \citenamefont {Zhang}, \citenamefont {Sun}, \citenamefont {Xu}, \citenamefont
  {Wang}, \citenamefont {Ji}, \citenamefont {Deng}, \citenamefont {Chen},
  \citenamefont {Liu},\ and\ \citenamefont {Pan}}]{Wu2016}%
  \BibitemOpen
  \bibfield  {author} {\bibinfo {author} {\bibfnamefont {Z.}~\bibnamefont
  {Wu}}, \bibinfo {author} {\bibfnamefont {L.}~\bibnamefont {Zhang}}, \bibinfo
  {author} {\bibfnamefont {W.}~\bibnamefont {Sun}}, \bibinfo {author}
  {\bibfnamefont {X.-T.}\ \bibnamefont {Xu}}, \bibinfo {author} {\bibfnamefont
  {B.-Z.}\ \bibnamefont {Wang}}, \bibinfo {author} {\bibfnamefont {S.-C.}\
  \bibnamefont {Ji}}, \bibinfo {author} {\bibfnamefont {Y.}~\bibnamefont
  {Deng}}, \bibinfo {author} {\bibfnamefont {S.}~\bibnamefont {Chen}}, \bibinfo
  {author} {\bibfnamefont {X.-J.}\ \bibnamefont {Liu}}, \ and\ \bibinfo
  {author} {\bibfnamefont {J.-W.}\ \bibnamefont {Pan}},\ }\href {\doibase
  10.1126/science.aaf6689} {\bibfield  {journal} {\bibinfo  {journal}
  {Science}\ }\textbf {\bibinfo {volume} {354}},\ \bibinfo {pages} {83}
  (\bibinfo {year} {2016})}\BibitemShut {NoStop}%
\bibitem [{\citenamefont {Hofstadter}(1976)}]{Hofstadter1976}%
  \BibitemOpen
  \bibfield  {author} {\bibinfo {author} {\bibfnamefont {D.~R.}\ \bibnamefont
  {Hofstadter}},\ }\href {\doibase 10.1103/PhysRevB.14.2239} {\bibfield
  {journal} {\bibinfo  {journal} {Phys. Rev. B}\ }\textbf {\bibinfo {volume}
  {14}},\ \bibinfo {pages} {2239} (\bibinfo {year} {1976})}\BibitemShut
  {NoStop}%
\bibitem [{\citenamefont {Goldman}\ \emph {et~al.}(2010)\citenamefont
  {Goldman}, \citenamefont {Satija}, \citenamefont {Nikolic}, \citenamefont
  {Bermudez}, \citenamefont {Martin-Delgado}, \citenamefont {Lewenstein},\ and\
  \citenamefont {Spielman}}]{Goldman2010}%
  \BibitemOpen
  \bibfield  {author} {\bibinfo {author} {\bibfnamefont {N.}~\bibnamefont
  {Goldman}}, \bibinfo {author} {\bibfnamefont {I.}~\bibnamefont {Satija}},
  \bibinfo {author} {\bibfnamefont {P.}~\bibnamefont {Nikolic}}, \bibinfo
  {author} {\bibfnamefont {A.}~\bibnamefont {Bermudez}}, \bibinfo {author}
  {\bibfnamefont {M.~A.}\ \bibnamefont {Martin-Delgado}}, \bibinfo {author}
  {\bibfnamefont {M.}~\bibnamefont {Lewenstein}}, \ and\ \bibinfo {author}
  {\bibfnamefont {I.~B.}\ \bibnamefont {Spielman}},\ }\href {\doibase
  10.1103/PhysRevLett.105.255302} {\bibfield  {journal} {\bibinfo  {journal}
  {Phys. Rev. Lett.}\ }\textbf {\bibinfo {volume} {105}},\ \bibinfo {pages}
  {255302} (\bibinfo {year} {2010})}\BibitemShut {NoStop}%
\bibitem [{\citenamefont {Dalibard}\ \emph {et~al.}(2011)\citenamefont
  {Dalibard}, \citenamefont {Gerbier}, \citenamefont {Juzeli\={u}nas},\ and\
  \citenamefont {{\"{O}}hberg}}]{Dalibard2011}%
  \BibitemOpen
  \bibfield  {author} {\bibinfo {author} {\bibfnamefont {J.}~\bibnamefont
  {Dalibard}}, \bibinfo {author} {\bibfnamefont {F.}~\bibnamefont {Gerbier}},
  \bibinfo {author} {\bibfnamefont {G.}~\bibnamefont {Juzeli\={u}nas}}, \ and\
  \bibinfo {author} {\bibfnamefont {P.}~\bibnamefont {{\"{O}}hberg}},\ }\href
  {\doibase 10.1103/RevModPhys.83.1523} {\bibfield  {journal} {\bibinfo
  {journal} {Rev. Mod. Phys.}\ }\textbf {\bibinfo {volume} {83}},\ \bibinfo
  {pages} {1523} (\bibinfo {year} {2011})}\BibitemShut {NoStop}%
\bibitem [{\citenamefont {Orth}\ \emph {et~al.}(2013)\citenamefont {Orth},
  \citenamefont {Cocks}, \citenamefont {Rachel}, \citenamefont {Buchhold},
  \citenamefont {Hur},\ and\ \citenamefont {Hofstetter}}]{Orth2013}%
  \BibitemOpen
  \bibfield  {author} {\bibinfo {author} {\bibfnamefont {P.~P.}\ \bibnamefont
  {Orth}}, \bibinfo {author} {\bibfnamefont {D.}~\bibnamefont {Cocks}},
  \bibinfo {author} {\bibfnamefont {S.}~\bibnamefont {Rachel}}, \bibinfo
  {author} {\bibfnamefont {M.}~\bibnamefont {Buchhold}}, \bibinfo {author}
  {\bibfnamefont {K.~L.}\ \bibnamefont {Hur}}, \ and\ \bibinfo {author}
  {\bibfnamefont {W.}~\bibnamefont {Hofstetter}},\ }\href@noop {} {\bibfield
  {journal} {\bibinfo  {journal} {J. Phys. B}\ }\textbf {\bibinfo {volume}
  {46}},\ \bibinfo {pages} {134004} (\bibinfo {year} {2013})}\BibitemShut
  {NoStop}%
\bibitem [{\citenamefont {Kane}\ and\ \citenamefont {Mele}(2005)}]{Kane2005}%
  \BibitemOpen
  \bibfield  {author} {\bibinfo {author} {\bibfnamefont {C.~L.}\ \bibnamefont
  {Kane}}\ and\ \bibinfo {author} {\bibfnamefont {E.~J.}\ \bibnamefont
  {Mele}},\ }\href@noop {} {\bibfield  {journal} {\bibinfo  {journal} {Phys.
  Rev. Lett.}\ }\textbf {\bibinfo {volume} {95}},\ \bibinfo {pages} {226801}
  (\bibinfo {year} {2005})}\BibitemShut {NoStop}%
\bibitem [{\citenamefont {Xu}\ and\ \citenamefont {Moore}(2006)}]{Xu2006}%
  \BibitemOpen
  \bibfield  {author} {\bibinfo {author} {\bibfnamefont {C.}~\bibnamefont
  {Xu}}\ and\ \bibinfo {author} {\bibfnamefont {J.~E.}\ \bibnamefont {Moore}},\
  }\href {\doibase 10.1103/PhysRevB.73.045322} {\bibfield  {journal} {\bibinfo
  {journal} {Phys. Rev. B}\ }\textbf {\bibinfo {volume} {73}},\ \bibinfo
  {pages} {045322} (\bibinfo {year} {2006})}\BibitemShut {NoStop}%
\bibitem [{\citenamefont {Thouless}\ \emph {et~al.}(1982)\citenamefont
  {Thouless}, \citenamefont {Kohmoto}, \citenamefont {Nightingale},\ and\
  \citenamefont {den Nijs}}]{Thouless1982}%
  \BibitemOpen
  \bibfield  {author} {\bibinfo {author} {\bibfnamefont {D.~J.}\ \bibnamefont
  {Thouless}}, \bibinfo {author} {\bibfnamefont {M.}~\bibnamefont {Kohmoto}},
  \bibinfo {author} {\bibfnamefont {M.~P.}\ \bibnamefont {Nightingale}}, \ and\
  \bibinfo {author} {\bibfnamefont {M.}~\bibnamefont {den Nijs}},\ }\href@noop
  {} {\bibfield  {journal} {\bibinfo  {journal} {Phys. Rev. Lett.}\ }\textbf
  {\bibinfo {volume} {49}},\ \bibinfo {pages} {405} (\bibinfo {year}
  {1982})}\BibitemShut {NoStop}%
\bibitem [{\citenamefont {Niu}\ \emph {et~al.}(1985)\citenamefont {Niu},
  \citenamefont {Thouless},\ and\ \citenamefont {Wu}}]{Niu1985}%
  \BibitemOpen
  \bibfield  {author} {\bibinfo {author} {\bibfnamefont {Q.}~\bibnamefont
  {Niu}}, \bibinfo {author} {\bibfnamefont {D.~J.}\ \bibnamefont {Thouless}}, \
  and\ \bibinfo {author} {\bibfnamefont {Y.~S.}\ \bibnamefont {Wu}},\ }\href
  {\doibase 10.1103/PhysRevB.31.3372} {\bibfield  {journal} {\bibinfo
  {journal} {Phys. Rev. B}\ }\textbf {\bibinfo {volume} {31}},\ \bibinfo
  {pages} {3372} (\bibinfo {year} {1985})}\BibitemShut {NoStop}%
\bibitem [{\citenamefont {Sheng}\ \emph {et~al.}(2006)\citenamefont {Sheng},
  \citenamefont {Weng}, \citenamefont {Sheng},\ and\ \citenamefont
  {Haldane}}]{Sheng2006}%
  \BibitemOpen
  \bibfield  {author} {\bibinfo {author} {\bibfnamefont {D.~N.}\ \bibnamefont
  {Sheng}}, \bibinfo {author} {\bibfnamefont {Z.~Y.}\ \bibnamefont {Weng}},
  \bibinfo {author} {\bibfnamefont {L.}~\bibnamefont {Sheng}}, \ and\ \bibinfo
  {author} {\bibfnamefont {F.~D.~M.}\ \bibnamefont {Haldane}},\ }\href@noop {}
  {\bibfield  {journal} {\bibinfo  {journal} {Phys. Rev. Lett.}\ }\textbf
  {\bibinfo {volume} {\textbf{97}}},\ \bibinfo {pages} {036808} (\bibinfo
  {year} {2006})}\BibitemShut {NoStop}%
\bibitem [{\citenamefont {Lee}\ and\ \citenamefont {Ryu}(2008)}]{Lee2008}%
  \BibitemOpen
  \bibfield  {author} {\bibinfo {author} {\bibfnamefont {S.~S.}\ \bibnamefont
  {Lee}}\ and\ \bibinfo {author} {\bibfnamefont {S.}~\bibnamefont {Ryu}},\
  }\href {\doibase 10.1103/PhysRevLett.100.186807} {\bibfield  {journal}
  {\bibinfo  {journal} {Phys. Rev. Lett.}\ }\textbf {\bibinfo {volume} {100}},\
  \bibinfo {pages} {186807} (\bibinfo {year} {2008})}\BibitemShut {NoStop}%
\bibitem [{\citenamefont {Ishikawa}\ and\ \citenamefont
  {Matsuyama}(1987)}]{Ishikawa1987}%
  \BibitemOpen
  \bibfield  {author} {\bibinfo {author} {\bibfnamefont {K.}~\bibnamefont
  {Ishikawa}}\ and\ \bibinfo {author} {\bibfnamefont {T.}~\bibnamefont
  {Matsuyama}},\ }\href {\doibase 10.1016/0550-3213(87)90160-X} {\bibfield
  {journal} {\bibinfo  {journal} {Nuclear Physics, Section B}\ }\textbf
  {\bibinfo {volume} {280}},\ \bibinfo {pages} {523} (\bibinfo {year}
  {1987})}\BibitemShut {NoStop}%
\bibitem [{\citenamefont {Wang}\ \emph {et~al.}(2010)\citenamefont {Wang},
  \citenamefont {Qi},\ and\ \citenamefont {Zhang}}]{Wang2010}%
  \BibitemOpen
  \bibfield  {author} {\bibinfo {author} {\bibfnamefont {Z.}~\bibnamefont
  {Wang}}, \bibinfo {author} {\bibfnamefont {X.~L.}\ \bibnamefont {Qi}}, \ and\
  \bibinfo {author} {\bibfnamefont {S.~C.}\ \bibnamefont {Zhang}},\ }\href
  {\doibase 10.1103/PhysRevLett.105.256803} {\bibfield  {journal} {\bibinfo
  {journal} {Physical Review Letters}\ }\textbf {\bibinfo {volume} {105}},\
  \bibinfo {pages} {256803} (\bibinfo {year} {2010})},\ \Eprint
  {http://arxiv.org/abs/1004.4229} {1004.4229} \BibitemShut {NoStop}%
\bibitem [{\citenamefont {Essin}\ and\ \citenamefont
  {Gurarie}(2015)}]{Essin2015}%
  \BibitemOpen
  \bibfield  {author} {\bibinfo {author} {\bibfnamefont {A.~M.}\ \bibnamefont
  {Essin}}\ and\ \bibinfo {author} {\bibfnamefont {V.}~\bibnamefont
  {Gurarie}},\ }\href {\doibase 10.1088/1751-8113/48/11/11FT01} {\bibfield
  {journal} {\bibinfo  {journal} {J. Phs. A-Math. Theor.}\ }\textbf {\bibinfo
  {volume} {48}},\ \bibinfo {pages} {11FT01} (\bibinfo {year} {2015})},\
  \Eprint {http://arxiv.org/abs/1410.2829} {1410.2829} \BibitemShut {NoStop}%
\bibitem [{\citenamefont {Prodan}\ \emph {et~al.}(2010)\citenamefont {Prodan},
  \citenamefont {Hughes},\ and\ \citenamefont {Bernevig}}]{Prodan2010}%
  \BibitemOpen
  \bibfield  {author} {\bibinfo {author} {\bibfnamefont {E.}~\bibnamefont
  {Prodan}}, \bibinfo {author} {\bibfnamefont {T.~L.}\ \bibnamefont {Hughes}},
  \ and\ \bibinfo {author} {\bibfnamefont {B.~A.}\ \bibnamefont {Bernevig}},\
  }\href {\doibase 10.1103/PhysRevLett.105.115501} {\bibfield  {journal}
  {\bibinfo  {journal} {Phys. Rev. Lett.}\ }\textbf {\bibinfo {volume} {105}},\
  \bibinfo {pages} {115501} (\bibinfo {year} {2010})}\BibitemShut {NoStop}%
\bibitem [{\citenamefont {Petrescu}\ \emph {et~al.}(2012)\citenamefont
  {Petrescu}, \citenamefont {Houck},\ and\ \citenamefont {{Le
  Hur}}}]{Petrescu2012}%
  \BibitemOpen
  \bibfield  {author} {\bibinfo {author} {\bibfnamefont {A.}~\bibnamefont
  {Petrescu}}, \bibinfo {author} {\bibfnamefont {A.~A.}\ \bibnamefont {Houck}},
  \ and\ \bibinfo {author} {\bibfnamefont {K.}~\bibnamefont {{Le Hur}}},\
  }\href {\doibase 10.1103/PhysRevA.86.053804} {\bibfield  {journal} {\bibinfo
  {journal} {Phys. Rev. A}\ }\textbf {\bibinfo {volume} {86}},\ \bibinfo
  {pages} {053804} (\bibinfo {year} {2012})}\BibitemShut {NoStop}%
\bibitem [{\citenamefont {Bianco}\ and\ \citenamefont
  {Resta}(2011)}]{Bianco2011}%
  \BibitemOpen
  \bibfield  {author} {\bibinfo {author} {\bibfnamefont {R.}~\bibnamefont
  {Bianco}}\ and\ \bibinfo {author} {\bibfnamefont {R.}~\bibnamefont {Resta}},\
  }\href {\doibase 10.1103/PhysRevB.84.241106} {\bibfield  {journal} {\bibinfo
  {journal} {Phys. Rev. B}\ }\textbf {\bibinfo {volume} {84}},\ \bibinfo
  {pages} {241106(R)} (\bibinfo {year} {2011})}\BibitemShut {NoStop}%
\bibitem [{\citenamefont {{Tran}}\ \emph {et~al.}(2015)\citenamefont {{Tran}},
  \citenamefont {{Dauphin}}, \citenamefont {{Goldman}},\ and\ \citenamefont
  {{Gaspard}}}]{Tran2015}%
  \BibitemOpen
  \bibfield  {author} {\bibinfo {author} {\bibfnamefont {D.-T.}\ \bibnamefont
  {{Tran}}}, \bibinfo {author} {\bibfnamefont {A.}~\bibnamefont {{Dauphin}}},
  \bibinfo {author} {\bibfnamefont {N.}~\bibnamefont {{Goldman}}}, \ and\
  \bibinfo {author} {\bibfnamefont {P.}~\bibnamefont {{Gaspard}}},\ }\href@noop
  {} {\bibfield  {journal} {\bibinfo  {journal} {Phys. Rev. B}\ }\textbf
  {\bibinfo {volume} {91}},\ \bibinfo {pages} {085125} (\bibinfo {year}
  {2015})}\BibitemShut {NoStop}%
\bibitem [{\citenamefont {Amaricci}\ \emph {et~al.}(2017)\citenamefont
  {Amaricci}, \citenamefont {Privitera}, \citenamefont {Petocchi},
  \citenamefont {Capone}, \citenamefont {Sangiovanni},\ and\ \citenamefont
  {Trauzettel}}]{Amaricci2017}%
  \BibitemOpen
  \bibfield  {author} {\bibinfo {author} {\bibfnamefont {A.}~\bibnamefont
  {Amaricci}}, \bibinfo {author} {\bibfnamefont {L.}~\bibnamefont {Privitera}},
  \bibinfo {author} {\bibfnamefont {F.}~\bibnamefont {Petocchi}}, \bibinfo
  {author} {\bibfnamefont {M.}~\bibnamefont {Capone}}, \bibinfo {author}
  {\bibfnamefont {G.}~\bibnamefont {Sangiovanni}}, \ and\ \bibinfo {author}
  {\bibfnamefont {B.}~\bibnamefont {Trauzettel}},\ }\href {\doibase
  10.1103/PhysRevB.95.205120} {\bibfield  {journal} {\bibinfo  {journal} {Phys.
  Rev. B}\ }\textbf {\bibinfo {volume} {95}},\ \bibinfo {pages} {205120}
  (\bibinfo {year} {2017})}\BibitemShut {NoStop}%
\bibitem [{\citenamefont {Irsigler}\ \emph
  {et~al.}(2019{\natexlab{a}})\citenamefont {Irsigler}, \citenamefont {Zheng},\
  and\ \citenamefont {Hofstetter}}]{Irsigler2019}%
  \BibitemOpen
  \bibfield  {author} {\bibinfo {author} {\bibfnamefont {B.}~\bibnamefont
  {Irsigler}}, \bibinfo {author} {\bibfnamefont {J.-H.}\ \bibnamefont {Zheng}},
  \ and\ \bibinfo {author} {\bibfnamefont {W.}~\bibnamefont {Hofstetter}},\
  }\href {\doibase 10.1103/PhysRevLett.122.010406} {\bibfield  {journal}
  {\bibinfo  {journal} {Phys. Rev. Lett.}\ }\textbf {\bibinfo {volume} {122}},\
  \bibinfo {pages} {010406} (\bibinfo {year} {2019}{\natexlab{a}})}\BibitemShut
  {NoStop}%
\bibitem [{\citenamefont {Marrazzo}\ and\ \citenamefont
  {Resta}(2017)}]{Marrazzo2017}%
  \BibitemOpen
  \bibfield  {author} {\bibinfo {author} {\bibfnamefont {A.}~\bibnamefont
  {Marrazzo}}\ and\ \bibinfo {author} {\bibfnamefont {R.}~\bibnamefont
  {Resta}},\ }\href {\doibase 10.1103/PhysRevB.95.121114} {\bibfield  {journal}
  {\bibinfo  {journal} {Phys. Rev. B}\ }\textbf {\bibinfo {volume} {95}},\
  \bibinfo {pages} {121114(R)} (\bibinfo {year} {2017})}\BibitemShut {NoStop}%
\bibitem [{\citenamefont {Irsigler}\ \emph
  {et~al.}(2019{\natexlab{b}})\citenamefont {Irsigler}, \citenamefont {Zheng},\
  and\ \citenamefont {Hofstetter}}]{Irsigler2019b}%
  \BibitemOpen
  \bibfield  {author} {\bibinfo {author} {\bibfnamefont {B.}~\bibnamefont
  {Irsigler}}, \bibinfo {author} {\bibfnamefont {J.-H.}\ \bibnamefont {Zheng}},
  \ and\ \bibinfo {author} {\bibfnamefont {W.}~\bibnamefont {Hofstetter}},\
  }\href {\doibase 10.1103/PhysRevA.100.023610} {\bibfield  {journal} {\bibinfo
   {journal} {Phys. Rev. A}\ }\textbf {\bibinfo {volume} {100}},\ \bibinfo
  {pages} {023610} (\bibinfo {year} {2019}{\natexlab{b}})}\BibitemShut
  {NoStop}%
\bibitem [{\citenamefont {Caio}\ \emph {et~al.}(2019)\citenamefont {Caio},
  \citenamefont {M\"{o}ller}, \citenamefont {Cooper},\ and\ \citenamefont
  {Bhaseen}}]{Caio2019}%
  \BibitemOpen
  \bibfield  {author} {\bibinfo {author} {\bibfnamefont {M.~D.}\ \bibnamefont
  {Caio}}, \bibinfo {author} {\bibfnamefont {G.}~\bibnamefont {M\"{o}ller}},
  \bibinfo {author} {\bibfnamefont {N.~R.}\ \bibnamefont {Cooper}}, \ and\
  \bibinfo {author} {\bibfnamefont {M.~J.}\ \bibnamefont {Bhaseen}},\ }\href
  {\doibase 10.1038/s41567-018-0390-7} {\bibfield  {journal} {\bibinfo
  {journal} {Nat. Phys.}\ }\textbf {\bibinfo {volume} {15}},\ \bibinfo {pages}
  {257} (\bibinfo {year} {2019})}\BibitemShut {NoStop}%
\bibitem [{\citenamefont {Schine}\ \emph {et~al.}(2019)\citenamefont {Schine},
  \citenamefont {Chalupnik}, \citenamefont {Can}, \citenamefont {Gromov},\ and\
  \citenamefont {Simon}}]{Schine2019}%
  \BibitemOpen
  \bibfield  {author} {\bibinfo {author} {\bibfnamefont {N.}~\bibnamefont
  {Schine}}, \bibinfo {author} {\bibfnamefont {M.}~\bibnamefont {Chalupnik}},
  \bibinfo {author} {\bibfnamefont {T.}~\bibnamefont {Can}}, \bibinfo {author}
  {\bibfnamefont {A.}~\bibnamefont {Gromov}}, \ and\ \bibinfo {author}
  {\bibfnamefont {J.}~\bibnamefont {Simon}},\ }\href {\doibase
  10.1038/s41586-018-0817-4} {\bibfield  {journal} {\bibinfo  {journal}
  {Nature}\ }\textbf {\bibinfo {volume} {565}},\ \bibinfo {pages} {173}
  (\bibinfo {year} {2019})}\BibitemShut {NoStop}%
\bibitem [{\citenamefont {Cocks}\ \emph {et~al.}(2012)\citenamefont {Cocks},
  \citenamefont {Orth}, \citenamefont {Rachel}, \citenamefont {Buchhold},
  \citenamefont {Le~Hur},\ and\ \citenamefont {Hofstetter}}]{Cocks2012}%
  \BibitemOpen
  \bibfield  {author} {\bibinfo {author} {\bibfnamefont {D.}~\bibnamefont
  {Cocks}}, \bibinfo {author} {\bibfnamefont {P.~P.}\ \bibnamefont {Orth}},
  \bibinfo {author} {\bibfnamefont {S.}~\bibnamefont {Rachel}}, \bibinfo
  {author} {\bibfnamefont {M.}~\bibnamefont {Buchhold}}, \bibinfo {author}
  {\bibfnamefont {K.}~\bibnamefont {Le~Hur}}, \ and\ \bibinfo {author}
  {\bibfnamefont {W.}~\bibnamefont {Hofstetter}},\ }\href@noop {} {\bibfield
  {journal} {\bibinfo  {journal} {Phys. Rev. Lett.}\ }\textbf {\bibinfo
  {volume} {109}},\ \bibinfo {pages} {205303} (\bibinfo {year}
  {2012})}\BibitemShut {NoStop}%
\bibitem [{\citenamefont {Kumar}\ \emph {et~al.}(2016)\citenamefont {Kumar},
  \citenamefont {Mertz},\ and\ \citenamefont {Hofstetter}}]{Kumar2016}%
  \BibitemOpen
  \bibfield  {author} {\bibinfo {author} {\bibfnamefont {P.}~\bibnamefont
  {Kumar}}, \bibinfo {author} {\bibfnamefont {T.}~\bibnamefont {Mertz}}, \ and\
  \bibinfo {author} {\bibfnamefont {W.}~\bibnamefont {Hofstetter}},\ }\href
  {\doibase 10.1103/PhysRevB.94.115161} {\bibfield  {journal} {\bibinfo
  {journal} {Phys. Rev. B}\ }\textbf {\bibinfo {volume} {94}},\ \bibinfo
  {pages} {115161} (\bibinfo {year} {2016})}\BibitemShut {NoStop}%
\bibitem [{\citenamefont {Irsigler}\ \emph
  {et~al.}(2019{\natexlab{c}})\citenamefont {Irsigler}, \citenamefont {Zheng},
  \citenamefont {Hafez-Torbati},\ and\ \citenamefont
  {Hofstetter}}]{Irsigler2019a}%
  \BibitemOpen
  \bibfield  {author} {\bibinfo {author} {\bibfnamefont {B.}~\bibnamefont
  {Irsigler}}, \bibinfo {author} {\bibfnamefont {J.-H.}\ \bibnamefont {Zheng}},
  \bibinfo {author} {\bibfnamefont {M.}~\bibnamefont {Hafez-Torbati}}, \ and\
  \bibinfo {author} {\bibfnamefont {W.}~\bibnamefont {Hofstetter}},\ }\href
  {\doibase 10.1103/PhysRevA.99.043628} {\bibfield  {journal} {\bibinfo
  {journal} {Phys. Rev. A}\ }\textbf {\bibinfo {volume} {99}},\ \bibinfo
  {pages} {043628} (\bibinfo {year} {2019}{\natexlab{c}})}\BibitemShut
  {NoStop}%
\bibitem [{\citenamefont {Georges}\ \emph {et~al.}(1996)\citenamefont
  {Georges}, \citenamefont {Kotliar}, \citenamefont {Krauth},\ and\
  \citenamefont {Rozenberg}}]{Georges1996}%
  \BibitemOpen
  \bibfield  {author} {\bibinfo {author} {\bibfnamefont {A.}~\bibnamefont
  {Georges}}, \bibinfo {author} {\bibfnamefont {G.}~\bibnamefont {Kotliar}},
  \bibinfo {author} {\bibfnamefont {W.}~\bibnamefont {Krauth}}, \ and\ \bibinfo
  {author} {\bibfnamefont {M.~J.}\ \bibnamefont {Rozenberg}},\ }\href {\doibase
  10.1103/RevModPhys.68.13} {\bibfield  {journal} {\bibinfo  {journal} {Rev.
  Mod. Phys.}\ }\textbf {\bibinfo {volume} {68}},\ \bibinfo {pages} {13}
  (\bibinfo {year} {1996})}\BibitemShut {NoStop}%
\bibitem [{\citenamefont {Okamoto}\ and\ \citenamefont
  {Millis}(2004)}]{Okamoto2004}%
  \BibitemOpen
  \bibfield  {author} {\bibinfo {author} {\bibfnamefont {S.}~\bibnamefont
  {Okamoto}}\ and\ \bibinfo {author} {\bibfnamefont {A.~J.}\ \bibnamefont
  {Millis}},\ }\href {\doibase 10.1103/PhysRevB.70.241104} {\bibfield
  {journal} {\bibinfo  {journal} {Phys. Rev. B}\ }\textbf {\bibinfo {volume}
  {70}},\ \bibinfo {pages} {241104(R)} (\bibinfo {year} {2004})}\BibitemShut
  {NoStop}%
\bibitem [{\citenamefont {Helmes}\ \emph {et~al.}(2008)\citenamefont {Helmes},
  \citenamefont {Costi},\ and\ \citenamefont {Rosch}}]{Helmes2008}%
  \BibitemOpen
  \bibfield  {author} {\bibinfo {author} {\bibfnamefont {R.~W.}\ \bibnamefont
  {Helmes}}, \bibinfo {author} {\bibfnamefont {T.~A.}\ \bibnamefont {Costi}}, \
  and\ \bibinfo {author} {\bibfnamefont {A.}~\bibnamefont {Rosch}},\ }\href
  {\doibase 10.1103/PhysRevLett.100.056403} {\bibfield  {journal} {\bibinfo
  {journal} {Phys. Rev. Lett.}\ }\textbf {\bibinfo {volume} {100}},\ \bibinfo
  {pages} {056403} (\bibinfo {year} {2008})}\BibitemShut {NoStop}%
\bibitem [{\citenamefont {Snoek}\ \emph {et~al.}(2008)\citenamefont {Snoek},
  \citenamefont {Titvinidze}, \citenamefont {T{\H{o}}ke}, \citenamefont
  {Byczuk},\ and\ \citenamefont {Hofstetter}}]{Snoek2008}%
  \BibitemOpen
  \bibfield  {author} {\bibinfo {author} {\bibfnamefont {M.}~\bibnamefont
  {Snoek}}, \bibinfo {author} {\bibfnamefont {I.}~\bibnamefont {Titvinidze}},
  \bibinfo {author} {\bibfnamefont {C.}~\bibnamefont {T{\H{o}}ke}}, \bibinfo
  {author} {\bibfnamefont {K.}~\bibnamefont {Byczuk}}, \ and\ \bibinfo {author}
  {\bibfnamefont {W.}~\bibnamefont {Hofstetter}},\ }\href {\doibase
  10.1088/1367-2630/10/9/093008} {\bibfield  {journal} {\bibinfo  {journal}
  {New J. Phys.}\ }\textbf {\bibinfo {volume} {10}},\ \bibinfo {pages} {093008}
  (\bibinfo {year} {2008})}\BibitemShut {NoStop}%
\bibitem [{\citenamefont {Gull}\ \emph {et~al.}(2011)\citenamefont {Gull},
  \citenamefont {Millis}, \citenamefont {Lichtenstein}, \citenamefont
  {Rubtsov}, \citenamefont {Troyer},\ and\ \citenamefont {Werner}}]{Gull2011}%
  \BibitemOpen
  \bibfield  {author} {\bibinfo {author} {\bibfnamefont {E.}~\bibnamefont
  {Gull}}, \bibinfo {author} {\bibfnamefont {A.~J.}\ \bibnamefont {Millis}},
  \bibinfo {author} {\bibfnamefont {A.~I.}\ \bibnamefont {Lichtenstein}},
  \bibinfo {author} {\bibfnamefont {A.~N.}\ \bibnamefont {Rubtsov}}, \bibinfo
  {author} {\bibfnamefont {M.}~\bibnamefont {Troyer}}, \ and\ \bibinfo {author}
  {\bibfnamefont {P.}~\bibnamefont {Werner}},\ }\href {\doibase
  10.1103/RevModPhys.83.349} {\bibfield  {journal} {\bibinfo  {journal} {Rev.
  Mod. Phys.}\ }\textbf {\bibinfo {volume} {83}},\ \bibinfo {pages} {349}
  (\bibinfo {year} {2011})}\BibitemShut {NoStop}%
\bibitem [{\citenamefont {Wang}\ and\ \citenamefont {Zhang}(2012)}]{Wang2012a}%
  \BibitemOpen
  \bibfield  {author} {\bibinfo {author} {\bibfnamefont {Z.}~\bibnamefont
  {Wang}}\ and\ \bibinfo {author} {\bibfnamefont {S.-C.}\ \bibnamefont
  {Zhang}},\ }\href {\doibase 10.1103/PhysRevX.2.031008} {\bibfield  {journal}
  {\bibinfo  {journal} {Phys. Rev. X}\ }\textbf {\bibinfo {volume} {2}},\
  \bibinfo {pages} {031008} (\bibinfo {year} {2012})}\BibitemShut {NoStop}%
\bibitem [{\citenamefont {Abanin}\ and\ \citenamefont
  {Pesin}(2012)}]{Abanin2012}%
  \BibitemOpen
  \bibfield  {author} {\bibinfo {author} {\bibfnamefont {D.~A.}\ \bibnamefont
  {Abanin}}\ and\ \bibinfo {author} {\bibfnamefont {D.~A.}\ \bibnamefont
  {Pesin}},\ }\href {\doibase 10.1103/PhysRevLett.109.066802} {\bibfield
  {journal} {\bibinfo  {journal} {Phys. Rev. Lett.}\ }\textbf {\bibinfo
  {volume} {109}},\ \bibinfo {pages} {066802} (\bibinfo {year}
  {2012})}\BibitemShut {NoStop}%
\bibitem [{\citenamefont {Vanhala}\ \emph {et~al.}(2016)\citenamefont
  {Vanhala}, \citenamefont {Siro}, \citenamefont {Liang}, \citenamefont
  {Troyer}, \citenamefont {Harju},\ and\ \citenamefont
  {T\"orm\"a}}]{Vanhala2016}%
  \BibitemOpen
  \bibfield  {author} {\bibinfo {author} {\bibfnamefont {T.~I.}\ \bibnamefont
  {Vanhala}}, \bibinfo {author} {\bibfnamefont {T.}~\bibnamefont {Siro}},
  \bibinfo {author} {\bibfnamefont {L.}~\bibnamefont {Liang}}, \bibinfo
  {author} {\bibfnamefont {M.}~\bibnamefont {Troyer}}, \bibinfo {author}
  {\bibfnamefont {A.}~\bibnamefont {Harju}}, \ and\ \bibinfo {author}
  {\bibfnamefont {P.}~\bibnamefont {T\"orm\"a}},\ }\href {\doibase
  10.1103/PhysRevLett.116.225305} {\bibfield  {journal} {\bibinfo  {journal}
  {Phys. Rev. Lett.}\ }\textbf {\bibinfo {volume} {116}},\ \bibinfo {pages}
  {225305} (\bibinfo {year} {2016})}\BibitemShut {NoStop}%
\bibitem [{\citenamefont {Zheng}\ \emph {et~al.}(2020)\citenamefont {Zheng},
  \citenamefont {Irsigler}, \citenamefont {Jiang}, \citenamefont {Weitenberg},\
  and\ \citenamefont {Hofstetter}}]{Zheng2019}%
  \BibitemOpen
  \bibfield  {author} {\bibinfo {author} {\bibfnamefont {J.-H.}\ \bibnamefont
  {Zheng}}, \bibinfo {author} {\bibfnamefont {B.}~\bibnamefont {Irsigler}},
  \bibinfo {author} {\bibfnamefont {L.}~\bibnamefont {Jiang}}, \bibinfo
  {author} {\bibfnamefont {C.}~\bibnamefont {Weitenberg}}, \ and\ \bibinfo
  {author} {\bibfnamefont {W.}~\bibnamefont {Hofstetter}},\ }\href {\doibase
  10.1103/PhysRevA.101.013631} {\bibfield  {journal} {\bibinfo  {journal}
  {Phys. Rev. A}\ }\textbf {\bibinfo {volume} {101}},\ \bibinfo {pages}
  {013631} (\bibinfo {year} {2020})}\BibitemShut {NoStop}%
\bibitem [{\citenamefont {Scheurer}\ \emph {et~al.}(2015)\citenamefont
  {Scheurer}, \citenamefont {Rachel},\ and\ \citenamefont
  {Orth}}]{Scheurer2015}%
  \BibitemOpen
  \bibfield  {author} {\bibinfo {author} {\bibfnamefont {M.~S.}\ \bibnamefont
  {Scheurer}}, \bibinfo {author} {\bibfnamefont {S.}~\bibnamefont {Rachel}}, \
  and\ \bibinfo {author} {\bibfnamefont {P.~P.}\ \bibnamefont {Orth}},\ }\href
  {http://dx.doi.org/10.1038/srep08386} {\bibfield  {journal} {\bibinfo
  {journal} {Sci. Rep.}\ }\textbf {\bibinfo {volume} {5}},\ \bibinfo {pages}
  {8386} (\bibinfo {year} {2015})}\BibitemShut {NoStop}%
\bibitem [{\citenamefont {Fu}\ \emph {et~al.}(2007)\citenamefont {Fu},
  \citenamefont {Kane},\ and\ \citenamefont {Mele}}]{Fu2007}%
  \BibitemOpen
  \bibfield  {author} {\bibinfo {author} {\bibfnamefont {L.}~\bibnamefont
  {Fu}}, \bibinfo {author} {\bibfnamefont {C.~L.}\ \bibnamefont {Kane}}, \ and\
  \bibinfo {author} {\bibfnamefont {E.~J.}\ \bibnamefont {Mele}},\ }\href
  {\doibase 10.1103/PhysRevLett.98.106803} {\bibfield  {journal} {\bibinfo
  {journal} {Phys. Rev. Lett.}\ }\textbf {\bibinfo {volume} {98}},\ \bibinfo
  {pages} {106803} (\bibinfo {year} {2007})}\BibitemShut {NoStop}%
\end{thebibliography}%
\end{document}